\documentclass[11pt]{article}
\pdfoutput=1

\usepackage{a4wide}
\usepackage{amsmath}

\usepackage{fourier} 
\usepackage[scaled=0.875]{helvet} 

\RequirePackage{amsmath,amssymb,amsxtra,amsthm}

\RequirePackage{mathrsfs}

\RequirePackage{setspace}
\RequirePackage{array}
\RequirePackage{booktabs}

\RequirePackage[pdftex,final]{graphicx}

\RequirePackage{colortbl}
\RequirePackage{xcolor}
\colorlet{titlerowcolor}{gray!15}

\usepackage{cite}

\definecolor{blue3}{RGB}{31,119,180}
\definecolor{red3}{RGB}{214,39,40}
\definecolor{orange3}{RGB}{255,127,14}
\definecolor{green3}{RGB}{44,160,44}

\usepackage{colortbl}
\definecolor{lightgreen}{cmyk}{0.2, 0, 0.2, 0.2}
\definecolor{lightgray}{cmyk}{0.1,0.2,0,0.1}
\definecolor{lightgray2}{cmyk}{0.1,0.1,0,0.1}

\usepackage{hyperref}
\hypersetup{colorlinks=true,linkcolor=red3,citecolor=green3,urlcolor=teal,pdfencoding=auto}

\numberwithin{equation}{section}
\numberwithin{table}{section}
\numberwithin{figure}{section}

\author{
	\begin{minipage}{1.00\linewidth}
		\vspace{1cm}
		\begin{center}
			\begin{small}
				\textbf{Giorgio Leone}
				\\ \vspace{1cm}
				{\em Dipartimento di Fisica, Universit\`a di Torino}
				\\
				{\em Via Pietro Giuria 1, I-10125 Torino}
			\end{small}
		\end{center}
		\vspace{1cm}
	\end{minipage}
}

\date{}

\title{\vspace{3cm}
	\begin{huge}{Tachyons and Misaligned Supersymmetry in Orientifold Vacua} 	
	\end{huge}
	\\ \vspace{.7cm}
}

\begin{document}
	
	\begin{titlepage}
		\maketitle
		\thispagestyle{empty}
		
		\vspace{-14cm}
		\begin{flushright}
			{\bf    \today}
		\end{flushright}

		\vspace{11cm}
		
		\begin{abstract}
		    We extend the analysis of Misaligned Supersymmetry to orientifold vacua. The lack of modular invariance in the Klein bottle, annulus and M\"obius strip amplitudes implies that the IR properties of the physical spectrum are related to the UV behaviour of closed strings freely propagating between D-branes and O-planes, and \emph{vice versa}. We thus build sector averaged sums $\langle d (n) \rangle$ associated to both representations of each Riemann surface. We find that the $\langle d (n) \rangle$'s associated to the loop channel control the coupling of closed-string tachyons to D-branes and/or O-planes, and therefore their vanishing is a necessary, but not sufficient, condition for stability in the closed sector. Similarly, the sector averaged sums associated to the tree-level amplitudes encode the presence of tachyons in the physical spectrum, and their vanishing is a sufficient, but not necessary, condition for classical stability. 
			Using this technology, it is difficult, if not impossible, to formulate necessary and sufficient criteria for the absence of tachyons since, in orientifold vacua, they can be removed from the spectrum by the orientifold projection. Although in these cases one would naively expect a cancellation between $\langle d (n)\rangle$'s associated to different amplitudes, this does not occur in practice. We provide several examples in ten and lower dimensions supporting our results. 
		\end{abstract}
		
		\vfill

		{\small
			\begin{itemize}
				\item[E-mail:] {\tt giorgio.leone@unito.it}
			\end{itemize}
		}

	\end{titlepage}

\setstretch{1.15}


{		\hypersetup{linkcolor=black}
	\tableofcontents	}

\newpage
	
	\section{Introduction}
	
	A compelling question when studying string theory vacua concerns their stability. Space-time supersymmetry provides an elegant solution to such a problem, since the GSO  projection \cite{Gliozzi:1976jf,Gliozzi:1976qd} eliminates the tachyon from the string spectrum and the vanishing of vacuum-to-vacuum amplitudes on a generic Riemann surface guarantees that the background geometry is not modified even at the quantum level. These models admit a low energy description in terms of supergravity which, in turn, can accommodate low-energy supersymmetry breaking via, for instance, gaugino condensation \cite{Nilles:2004zg,Derendinger:1985kk,Dine:1985rz, Ferrara:1982qs}. Although this framework could give rise to an interesting phenomenology, it is intrinsically non-perturbative and thus its effects are difficult to control. Furthermore, the lack of experimental evidence for low-energy supersymmetry, the vast and diversified Landscape of string theory vacua and naturalness issues implied by Swampland considerations \cite{Cribiori:2021gbf, Coudarchet:2021qwc, DallAgata:2021nnr,Basile:2022vft} suggest that other scenarios, where supersymmetry is broken at higher energies, and possibly already at the string scale, are worth to be considered. In these cases, the presence of tachyons in the string spectrum or the emergence of tree-level or higher-loop tadpoles jeopardises the stability of classical vacua. Although the effects of tadpoles have been extensively investigated within the effective (super)gravity description \cite{Baykara:2022cwj,Basile:2018irz,Basile:2020mpt,Basile:2021krk,Raucci:2022bjw,Raucci:2022jgw,Antonelli:2019nar, Mourad:2016xbk,Mourad:2021qwf,Mourad:2021roa,Mourad:2022loy, Dudas:2002dg,Dudas:2010gi, Dudas:2000ff, Gubser:2001zr,DeWolfe:2001nz}, a full-fledged stringy analysis is still lacking since, to date, it is not known how to implement the Fischler-Susskind mechanism \cite{Fischler:1986ci,Fischler:1986tb} or simply describe the theory in the wrong vacuum \cite{Dudas:2004nd, Kitazawa:2008hv}. Much worse is the control of tachyonic vacua since the background is clearly unstable already at the classical level, and one has no knowledge of the off-shell tachyon potential. As a result, it is difficult, if not impossible, to find the real vacuum upon which the theory is built, and thus the perturbative analysis is no longer predictive. This is indeed the general situation for the condensation of the closed-string tachyon\footnote{Indeed, very few results have been obtained for the condensation of closed-string tachyons in very special set-ups \cite{Kaidi:2020jla,Hellerman:2007zz}}, while  the dynamics of a brane-antibrane system is under control in simple scenarios \cite{Sen:1998sm,Sen:2002in,Sen:2004nf, Sen:1999nx,Schnabl:2005gv}. 
	
It is thus important to find possible correlations between the presence/absence of physical tachyons and the properties of the {\em complete} spectrum of the string vacuum, including its massive excitations which indeed play an important role for the UV finiteness of String Theory.

For oriented closed strings this question was addressed long ago in \cite{Kutasov:1990sv} and few years later in \cite{Dienes:1994np} where it was observed that the massive spectrum of tachyon-free vacua actually enjoys a {\em misaligned supersymmetry} reflected in a slower exponential growth of the effective number of states, which translates into an oscillatory pattern of bosons and fermions\footnote{It is curious to see that the oscillations were first observed in \cite{Cudell:1992bi} in the QCD spectrum where it was argued that the alternating behaviour of baryonic and mesonic degrees of freedom is dictated by modular invariance.}. Using properties of the Rankin-Selberg-Zagier transform \cite{Rank, Selb, Zagier2}, a rigorous mathematical proof of the shape of such oscillations in classically stable vacua was then given in \cite{Angelantonj:2010ic}, where it was shown that their frequencies are set by the non-trivial zeroes of the Riemann zeta function. A thorough understanding of the implications of the absence of physical tachyons was actually gained in \cite{AFL} where, generalising the analysis of \cite{Cribiori1,Cribiori2} to any closed-string construction, it was proven that classical stability is actually linked to the vanishing of the {\em sector averaged sum} $\langle d(n) \rangle$ encoding the global large mass behaviour of physical degrees of freedom, as conjectured in \cite{Dienes:1994np}. Furthermore it was understood that the presence of space-time fermions is the real cause of the oscillations observed in the string spectrum, which manifest themselves also in tachyonic vacua \cite{AFL, Sara,Faraggi:2020fwg} and thus are not a sufficient condition for classical stability, as often stated in the literature. 

In the case of open strings and orientifold vacua a similar analysis is still lacking, since modular invariance is no longer present. As a result, there is no direct connection between light and massive states and, for instance, the large mass behaviour of the open-string spectrum does not give information about the presence or absence of open-string tachyons.  In fact, modular transformations connect different descriptions of the same amplitude, whereby loop diagrams are mapped to tree level propagation of closed strings between D-branes and/or orientifold planes. Therefore, one can at best relate IR properties of loop amplitudes to UV behaviour of tree-level ones, and {\em vice versa}. A partial understanding of this connection was indeed achieved in \cite{Niarchos1, Niarchos2} where it was shown that the large-mass behaviour of the sector averaged sum associated to the open-string spectrum governs the coupling of closed-string tachyons to D-branes. The purpose of the present paper is to extend the analysis of \cite{Niarchos1, Niarchos2} to more general set-ups, involving also unoriented strings, and to investigate {\em dual} conditions relating the absence of tachyons in the loop amplitudes to a similar sector averaged sum of closed strings propagating in the tree-level diagrams. Using the exact Rademacher expansion of characters, we show that the absence/decoupling of tachyons in the loop/tree-level amplitude implies the vanishing of the associated $\langle d (n) \rangle$, thus suggesting in a way the presence of misaligned supersymmetry also for open/unoriented strings. This behaviour reproduces what was found in the case of closed strings \cite{Dienes:1994np, Cribiori1, AFL}, although the interpretation is clearly different. Our analysis is based on general properties of the CFT on Riemann surfaces with vanishing Euler character, with boundaries and/or cross-caps, and it applies to many examples in ten and lower dimensions where supersymmetry is either broken directly at the string scale or spontaneously via a Scherk-Schwarz reduction or its T-dual M-theory breaking. The vanishing of the sector averaged sum in the open-string sector of the Sugimoto \cite{Sugimoto:1999tx} and type $0'\text{B}$ \cite{Sagnotti1, Sagnotti2,BS, AS} models already appeared in \cite{Cribiori1,Cribiori2}, where, however, there is no explicit discussion about the connection between the sector averaged sum and the role played by tachyons.  
 
Although such analysis connects the vanishing of $\langle d(n)\rangle$ of a given one-loop/tree-level amplitude with the absence of tachyons in the corresponding tree-level/one-loop counterpart, it fails in explaining the classical stability of vacua where tachyons are eliminated by the orientifold projection as, for instance, in the type $0'\text{B}$ theory. As a result, the real reason behind the absence of tachyons in these vacua is still elusive and requires new tools and ideas.

The paper is organised as follows: in Section \ref{Exact} the connection between the sector averaged sum associated to genus one amplitudes and the presence of open and closed tachyonic instabilities is described in full generality. We apply these results to some $10 d$ vacua in Section \ref{10dmodels} and to $9d$ orientifold constructions obtained through a Scherk-Schwarz and an M-theory breaking compactification in Sections \ref{Scherk-Schwarz orientifolds} and \ref{M-theory orientifolds}, respectively. The impossibility of explaining classical stability of orientifold vacua via the cancellations of $\langle d(n) \rangle$'s associated to different Riemann surfaces is discussed in Section \ref{discussion}. Section \ref{conclusion} summarises our results. Finally, the two appendices \ref{rational Scherk-Schwarz} and \ref{rational M-theory} collect some useful details on characters and vacuum amplitudes associated to $9d$ reductions with Scherk-Schwarz and M-theory breaking, respectively. 
	
	\section{UV-IR mixing and Misaligned Supersymmetry } \label{Exact}
	
	The interplay between the UV and IR regimes is a well known property of conformal field theories on Riemann surfaces with vanishing Euler characteristic. For closed strings on the torus this relation is guaranteed by modular invariance, which connects the light excitations with the asymptotic behaviour of massive states, a link which is at the heart of misaligned supersymmetry \cite{Dienes:1994np}. This should be contrasted to the case of world-sheets with the topology of the Klein bottle, the annulus and of the M\"obius strip. These surfaces can be built imposing anti-holomorphic involutions on their double covering tori, which trivialise the modular group. As a result, modular transformations provide different descriptions of the same amplitude. For instance, an $S$ transformation on the annulus amplitude maps the loop propagation of open strings to the tree level propagation of closed strings and {\em vice versa}, so that the UV behaviour of one description is linked to the IR behaviour of the dual interpretation\footnote{See, for instance, \cite{AS} for the construction of the Klein bottle, annulus and M\"obius strip surfaces from the double covering torus and the relation between loop and tree-level channels}. This makes the study of misaligned supersymmetry richer although less powerful in determining conditions for the absence of physical tachyons in orientifold vacua.

To introduce the stage, we shall briefly review misaligned supersymmetry for oriented closed strings following the analysis of \cite{AFL}, before turning to world-sheets with boundaries and cross-caps.  
	
	\subsection{A quick review of misaligned supersymmetry for closed strings}
	
	In the context of closed strings the behaviour of the degrees of freedom for large mass  can be used to investigate the classical stability \footnote{As explained in the introduction, {\em classical stability} is synonymous for the absence of tachyons in the tree level spectrum.} of the vacuum \cite{Cribiori1,Cribiori2,AFL,Dienes:1994np}. This result relies on two crucial points: modular invariance of the torus amplitude encoding the spectrum of the closed string and an exact formula for the growth of degrees of freedom at large mass obtained via the circle method developed in \cite{Hardy, Rademacher:1937a, Rademacher:1937b, Rademacher:1938, Rademacher:1938b }. Indeed, starting from the torus partition function 
		\begin{equation}
				{\mathscr T} = \sum_{a,b} {\mathscr N}_{ab}\, \chi_a \bar{\chi}_b 
				= \sum_{n,m} \sum_{a,b} {\mathscr N}_{ab}\, d_{ab} (n,m)\, q^{n} \bar q^m \, ,
	\end{equation}	
for $d$ uncompactified dimensions, written in terms of $M$ pseudo characters $\chi_a=\check{\chi}_a/\eta^{d-2}$ \cite{AFL,Vafa}, the {\em sector averaged sum} is defined as \cite{Dienes:1994np, Cribiori1,Cribiori2,AFL}
    \begin{equation} \label{sas}
    	\langle d (n) \rangle \left ( {\mathscr T}  \right )  =\sum_{a,b} {\mathscr N}_{ab}\, \Phi_{ab} (n)\, ,
    \end{equation}
where we have introduced the enveloping functions $\Phi_{ab}(n)$ to be described shortly. In $\chi_a$, the Dedekind eta function refers to the contribution of the world-sheet bosons associated to the uncompactified dimensions, while $\check{\chi}_a$ are the characters associated to the RCFT describing the world-sheet fermions and the compact bosons. Starting from the $q$-expansion
    \begin{equation} \label{qexp}
    	\chi_a(q)= \sum_{n \in \mathbb{Z}} d_a(n) q^{n+H_a} \,, \qquad \qquad \qquad H_a=h_a -c/24 \ ,
    \end{equation}
    a direct application of the  {\em Circle Method } \cite{Hardy, Rademacher:1937a, Rademacher:1937b, Rademacher:1938, Rademacher:1938b, Sussman2017} gives an exact expression for the degeneracies \cite{Manschot:2007ha, Dijkgraaf:2000fq, AFL}
    \begin{equation}
    	d_a (n) = \sum_{b \ | \ H_b <0} \sum_{\ell=1}^{\infty} \, Q^{ (\ell, n)}_{ab} \, f_b (\ell,n ) \, , \label{degch}
    \end{equation}
where the sum is restricted to the tachyonic characters with $H_b <0$.  This result depends on  the modified Bessel function of the first kind  $I_{\nu}(x)$ via
    \begin{equation}\label{bessel}
    	f_{b} (\ell , n) = \frac{2\pi d_b (0)}{\ell}\, \left(\frac{|H_b|}{n+H_a}\right)^{d/4} \, I_{d/2} \left( \frac{4\pi}{\ell} \sqrt{ |H_b | (n+ H_a )}\right)\,,
    \end{equation}
     and on the so called {\em generalised Kloosterman sum}
    \begin{equation}
    	Q^{(\ell,n)}_{ab} =  i^{1-d/2}\, \sum_{p=0\atop  (\ell,p)=1}^{\ell-1} \left( M^{-1}_{\ell,p} \right)_{ab} \, e^{-\frac{2\pi i}{\ell} (p  n_a - p' H_b )} \ . \label{Qfunct}
    \end{equation}
The matrices  $M_{\ell,p}$ represent the action of the modular transformation
    \begin{equation} \label{gamma}
    	\gamma_{p,\ell} = \begin{pmatrix} -p' & \frac{1+pp'}{\ell}\\-\ell & p\end{pmatrix} \ , \qquad \qquad \qquad \text{with $p' \ | \ $ } \gamma_{\ell,p}\in\text{SL}(2,\mathbb{Z}) \ ,
    \end{equation}
on the characters, as required by the Circle Method. Now, following \cite{AFL,Cribiori1,Cribiori2}, one can use the periodicity of the {\em generalised Kloosterman sums} 
\begin{equation}
Q^{(\ell,n+\ell )}_{ab} = Q^{(\ell,n)}_{ab}
\ ,
\end{equation}
to perform a further refinement of the spectrum, so that we select string states with fixed values of $Q$,
    \begin{equation}
    	\Phi_a (n,w ) = \sum_{b \ | \ H_b <0} \, Q^{ (\ell, w)}_{ab} \, f_b (\ell,n ) \ , \label{phidegch}
    \end{equation}
and allows us to introduce the enveloping functions
    \begin{equation}
    	\Phi_{ab}(n)= \sum_{\ell=1}^{\infty} \sum_{w=0}^{\ell-1} 	\Phi_a (n,w ) \overline \Phi_b (n,w+ H_a-\overline H_b ) \ ,
    \end{equation}
depending on the continuous variable $n$. This leads to the final expression for the sector averaged sum
     \begin{equation} \label{largemasssas}
   	\langle d(n) \rangle \sim \sum_{a,b \atop H_b = \bar H_a <0} {\mathscr N}_{ab}\,  d_b (0) \, \bar d _a (0)\, \frac{|H_b|^{(d-1)/2}}{2\, n^{(d+1)/2}}\, 
   	\sum_{\ell=1}^{\infty} \, \varphi (\ell) \, e^{\frac{8\pi}{\ell} \sqrt{|H_b|\, n}} \,,
   \end{equation} 
   which involves only level-matched tachyonic characters with $H_b <0$. Although the individual functions $\Phi_{ab}(n)$ exhibit an exponential growth dictated by the well known Hardy-Ramanujan estimate $O(e^{4 \pi \left ( \sqrt{c_L/24} + \sqrt{c_R/24} \right )   \sqrt{n}})$ \cite{Hardy}, $\langle d(n) \rangle $ can experience a milder behaviour $O (e^{ C_\text{eff} \sqrt{n}})$, dictated by the lowest value of $H_b $ allowed in the physical spectrum
   \begin{equation}
   	C_\text{eff} = 8 \pi \sqrt{|H_{b_\text{low}} |} \le 4 \pi \left ( \sqrt{c_L/24} + \sqrt{c_R/24} \right ) = C_\text{tot}\,. \label{deepestCeff}
   \end{equation}
    Indeed, thanks to the presence of the GSO matrix ${\mathscr N}_{ab}$,  $C_{\text{eff}}<C_{\text{tot}}$ holds whenever fermions are part of the spectrum, even in the presence of physical tachyons, whereas $C_{\text{eff}}=0$, and thus $\langle d(n) \rangle =0$, is valid if and only if the vacuum is classically stable. For our purpose, it is essential to stress that the result described in \eqref{largemasssas} relies {\em deeply} on the presence of modular invariance, which allows to characterise the vanishing of the sector averaged sum as the  {\em necessary and sufficient} condition for the absence of physical tachyons. In fact,  the generalised Kloosterman sums depend on the representation $M_{\ell ,p}$ of the modular transformations on the characters, for instance \cite{Vafa}
     \begin{equation} \label{Qmod}
     	\begin{aligned}
     		& Q_{ab}^{(1,n)} = i^{1-d/2}\, S_{ab} \ ,
     		\\
     		\\
     		& Q_{ab}^{(2,n)} = i^{1-d/2}\, (-1)^n\, \left (T^{\frac12}S T^2 S T^{\frac12} \right )_{ab}\ .
     	\end{aligned}
     \end{equation}
      The modular invariance of the GSO matrix then absorbs the $M_{\ell ,p}$'s from both the holomorphic and anti-holomorphic sectors, leaving just a dependence on those tachyonic characters that are level matched \cite{AFL}.
      
     \subsection{The orientifolded sector averaged sum} 
   
	For orientifold vacua \cite{Sagnotti:1987tw, Pradisi:1988xd, Horava:1989vt, BS, Bianchi:1990tb, Bianchi:1991eu} the conclusion reached in the previous section cannot hold since modular invariance is no longer a property of the vacuum-to-vacuum amplitudes. To compute the unoriented closed string spectrum, string states are projected to left-right invariant combinations through the world-sheet parity operator $\Omega$, resulting into the introduction of an involutive action on the double covering torus. The resulting Riemann surface  has then the topology of  the Klein bottle, $\mathcal{K}$. This surface is unoriented and involves the presence of a pair of cross-caps, which correspond to orientifold planes carrying a negative tension and a negative charge for RR fields \cite{Polchinski}. The consistency of the construction, related to the cancellation of RR tadpoles, requires the introduction of D-branes on which strings can end, thus describing the open sector of the vacuum. D-branes then imply that Riemann surfaces with boundaries be present in the perturbative expansion. As a result, aside from the torus and the Klein bottle, the relevant genus-one amplitudes are the Annulus, $\mathcal{A}$, with two boundaries, and the M\"obius strip, $\mathcal{M}$, with one boundary and one cross-cap, so that the combination $(\mathcal{A}+\mathcal{M})/2$ enforces the orientifold projection on the open-string spectrum. As anticipated, the surfaces $\mathcal{K}, \mathcal{A}$ and $ \mathcal{M}$ are built  through an anti-holomorphic involution of the double-covering torus, which is incompatible with modular invariance \cite{Bianchi:1989du}. 
	The action of  $\Omega$  on closed strings, and the nature of the open strings, imply that these amplitudes depend linearly on the characters, and can be conveniently written as 
	\begin{equation} \label{ampld}
		\mathcal{Z}_I=  \ \sum_{a=0}^{M-1} \text{Z}_I^a \ \chi_a = \sum_{a=0}^{M-1} \text{Z}_I^a \sum_{n=0}^{\infty} d_a(n) q^{n+H_a} \,,
	\end{equation}  
with $I=1,2,3$ labels  the Klein bottle, Annulus and Moebius strip amplitudes, respectively, while the $\text{Z}_I^a$ are suitable integers. In \eqref{ampld} $\chi_a$ are the {\em real} characters depending on the modulus $\tau$ of the double covering torus whose real part is fixed. For $\mathcal{K}$ and $\mathcal{A}$ the Teichm\"uller parameter is purely imaginary ($\tau= 2 i \tau_2$ and $\tau = \frac{i\tau_2}{2}$, respectively) so that the characters defined in \eqref{qexp} are real, whereas for the M\"obius strip surface the modulus of the covering torus is  $\frac{1}{2} + i \frac{\tau_2}{2}$, and one needs to introduce the extra phase $e^{-i \pi H_a}$ to make them real \cite{BS}. This extra phase will not play any role in the following discussion and thus, with an abuse of notation, we shall omit it. These amplitudes written in terms of the modulus $\tau$ all describe the one loop propagation of closed/open strings which, in $\mathcal{K}$ and $\mathcal{M}$ flip their orientation. It is conventional to refer to this description as the {\em loop-channel} which involves a {\em vertical proper-time} \cite{BS,AS}. 
The action of the modular group provides alternative physical descriptions of these amplitudes. In particular, the $S$ transformation\footnote{Actually, for the M\"obius strip this inversion of proper time is realised via the $P=T^{1/2} S T^2 S T^{1/2}$ transformation \cite{BS}.} exchanges the length $\sigma^1$ of the string with the proper time $\sigma^0$ on the world-sheet, so that the proper time is now {\em horizontal} and the resulting {\em tree-level channel} amplitudes 
	\begin{equation} \label{ampltr}
		\Tilde{\mathcal{Z}}_I=  \sum_{a=0}^{M-1} \Tilde{\text{Z}}_I^a \ \chi_a= \sum_{a=0}^{M-1} \Tilde{\text{Z}}_I^a \sum_{n=0}^{\infty} d_a(n) q^{n+H_a}  \,,
	\end{equation}   
describes the free propagation of closed strings between boundaries ({\em i.e.} D-branes) and cross-caps ({\em i.e.} O-planes) \cite{BS,AS}.  
Here $q=e^{-2 \pi \ell} $ is written in terms of the new horizontal proper time $\ell$, and for the M\"obius amplitudes it carries an extra minus sign, which originates from the non-vanishing real part of $\tau$. The $\Tilde{\text{Z}}_I^a$ can be obtained by the corresponding $\text{Z}_I^a$ once the $i^{1-d/2}S$ or $i^{1-d/2}P$ transformation is performed \footnote{The extra phases are present since, from our definitions, modular transformations are performed on pseudo-characters.}. 

   From the expressions \eqref{ampld} and \eqref{ampltr} it is straightforward to extract the large mass behaviour in the two channels which is encoded in the {\em sector averaged sums} associated to the corresponding amplitudes
	\begin{equation} \label{sasor}
		\left \langle d(n) \right \rangle  \left ( \mathcal{Z}_I \right ) =  \sum_{a=0}^{M-1} \ \text{Z}_I^a \ \Phi_a(n) \ , \qquad 	\left \langle d(n) \right \rangle \left ( \Tilde{\mathcal{Z}}_I \right ) =  \sum_{a=0}^{M-1} \ \Tilde{\text{Z}}_I^a \ \Phi_a(n ) \ .
	\end{equation}   
As in the closed-string case, 
\begin{equation}
\Phi_a (n) = \sum_{\ell =1}^\infty \sum_{w=0}^{\ell -1} \, \Phi_a (n,w )\ ,
\end{equation}
are the enveloping functions associated to the degeneracies of the real characters, once the continuation of $n$ to the reals and the refinement on the spectrum are employed. For the Klein bottle and Annulus amplitudes, the $\Phi_a (n, w)$ are simply given by eq. \eqref{phidegch} whereas in the Moebius strip amplitude they involve the new Kloosterman sums
   \begin{equation} \label{hattedkloost}
   	\hat{Q}_{ab}^{( \ell,w)}= (-1)^w Q^{(\ell,w)}_{a b}\,,
   \end{equation}   
where the alternating sign is clearly associated to the fixed real part of the Teichm\"uller parameter. The presence of $(-1)^w$ affects their periodicity in $w$, which is now $\hat \ell= \text{lcm}(2,\ell)$.

The properties of the Kloosterman sum drastically simplify the expressions of the enveloping functions. In fact as shown in \cite{Cribiori1,Cribiori2,AFL}, 
    \begin{equation}\label{Kloostann}
    	\begin{split}
    		\sum_{w=0}^{ \ell-1}  Q^{( \ell , w )}_{ab} &= 	\sum_{w=0}^{ \ell-1} \sum_{ p=0\atop ( p, \ell )=1}^{ \ell -1} e^{-\frac{2 \pi i}{ \ell} ( p w +  p '  H_b )}\, \left( M^{-1}_{ \ell , p}\right)_{ab} \ ,
    		\\
    		&= \sum_{ p=0\atop ( p, \ell )=1}^{ \ell -1}  e^{-\frac{2 \pi i}{ \ell}   p '  H_b }\, \left( M^{-1}_{ \ell , p}\right)_{ab}	\sum_{w=0}^{ \ell-1} e^{-\frac{2 \pi i}{ \ell}  p w} \  ,
    		\\
    		&= \sum_{ p=0\atop ( p, \ell )=1}^{ \ell -1}  e^{-\frac{2 \pi i}{ \ell}   p '  H_b }\, \left( M^{-1}_{ \ell , p}\right)_{ab} \ell \delta_{p,0} \, ,
    	\end{split}
    \end{equation}
which vanishes identically unless $\ell=1$, since only in this case $p=0$ is coprime with $\ell$. Therefore, the only contribution to the growth of the sector averaged sum comes from $\ell=1$, 
\begin{equation}\label{leadPhiKA}
\Phi_a (n) = \sum_{b \ | \ H_b <0} \, Q^{ (1, 0)}_{ab} \, f_b (1,n ) =   i^{1-d/2}\, \sum_{b \ | \ H_b <0} S_{ab} \, f_b (1,n)\,,
\end{equation}
and, in the last equality, we have used the explicit formula \eqref{Qmod} for $Q^{ (1, 0)}_{ab}$. Notice that this expression only involves the $S$ modular transformation that relates the $1$-loop amplitudes, $\mathcal{K}$ and $\mathcal{A}$, to their dual counterparts,  $\Tilde{\mathcal{K}}$ and $\Tilde{\mathcal{A}}$, describing the tree-level propagation of closed strings between pairs of O-planes and D-branes.

The M\"obius strip amplitude depends on the modified Kloosterman sum, which are also vanishing unless $\ell=2$. In fact,
   \begin{equation}\label{Kloostmob}
    	\begin{split}
    		\sum_{w=0}^{ \hat \ell-1}  \hat Q^{( \ell , w )}_{ab} &= \sum_{w=0}^{\hat\ell -1} (-1)^w \, \sum_{p=0 \atop (p,\ell)=1}^{\ell-1}\, e^{- \frac{2\pi i}{\ell} (pw + p' H_b )}\,  \left( M^{-1}_{ \ell , p}\right)_{ab}
		\\
		&=\sum_{ p=0\atop ( p, \ell )=1}^{ \ell -1}  e^{-\frac{2 \pi i}{ \ell}   p '  H_b }\, \left( M^{-1}_{ \ell , p}\right)_{ab}	\sum_{r=0}^{\frac{\hat \ell}{\text{gcd}(\hat \ell,\ell)}-1} \sum_{k=0}^{ \ell-1} e^{-\frac{2 \pi i}{ \ell} ( p- \frac{\ell}{2}) (k+ \ell r)} \ 
    		\\
    		&= \sum_{ p=0\atop ( p, \ell )=1}^{ \ell -1}  e^{-\frac{2 \pi i}{ \ell}   p '  H_b }\, \left( M^{-1}_{ \ell , p}\right)_{ab} \sum_{r=0}^{\frac{\hat \ell}{\text{gcd}(\hat \ell,\ell)}-1}  e^{-2 \pi i  ( p- \frac{\ell}{2})  r} \ell \, \delta_{p,\frac{\ell}{2}} \ ,
    	\end{split}
    \end{equation}
where in the second line we have written $w=k+\ell r$ and converted the sum over $w$ into the sums over $k$ and $r$. Clearly, the only non-vanishing term is for $\ell=2$ and $p=1$, so that
\begin{equation} \label{leadPhiM}
\sum_{w=0}^{1}\hat{Q}_{ab}^{(2,w)} = 2 i^{1-d/2} \  \left (T^{\frac12} S T^2 S T^{\frac12} \right )_{ab} = 2 i^{1-d/2}\, P_{ab} \,,
\end{equation}
and we recognise the modular transformation $P$ which connects the dual amplitudes $\mathcal{M}$ and $\Tilde{ \mathcal{M}}$, describing the orientifold projection of the open spectrum and the propagation of closed states between a boundary and a cross-cap, respectively.

We are now in the position to state the main result of our paper, which extends misaligned supersymmetry of oriented closed strings to the case of orientifold constructions. In fact, eq. \eqref{leadPhiKA}  implies
\begin{equation} 
	\left \langle d(n) \right \rangle \left (\mathcal{Z}_I \right )  = i^{1-d/2}  \ \sum_{a=0}^{M-1} \sum_{b \, |\, H_b <0} Z_{I}^{a} \, S_{ab} \, f_b (1 , n ) =   \sum_{b \,|\, H_b <0} \ \Tilde{\text{Z}}_I^b \   f_b(1, n) \ ,
\end{equation}
for the Klein bottle and annulus amplitudes, $I=1,2$, while, from eq. \eqref{leadPhiM}, it follows that
\begin{equation} 
	\left \langle d(n) \right \rangle \left (\mathcal{M} \right )  =2 i^{1-d/2}  \ \sum_{a=0}^{M-1} \sum_{b \, |\, H_b <0} Z^a_3 \, P_{ab} \, f_b (2 , n ) = 2  \sum_{b \,|\, H_b <0} \ \Tilde{\text{Z}}_3^b \   f_b(2, n) \ ,
\end{equation}
for the M\"obius strip amplitude. In both cases, the sector average sum associated to the loop-channel amplitudes grows exponentially if and only if a closed-string tachyons propagates in the tree-level channel. Using the explicit expression \eqref{bessel}, one can write
\begin{equation}\label{dsas}
	\left \langle d(n) \right \rangle \left (\mathcal{Z}_I \right ) =  \sqrt{\tfrac{\ell_I }{2}}  \sum_{b\, |\, H_b <0} \ \Tilde{\text{Z}}_I^b  \, d_b(0)  \, \frac{\left | H_b \right |^{(d-1)/4}}{n^{(d+1)/4}} \, e^{ \frac{4\pi}{\ell_I} \, \sqrt{\left | H_b \right | \, n}} \ ,
\end{equation}
with $\ell_I =1$ ($\ell_I =2$) for $I= 1, 2$ ($I=3$). Similarly, the sector averaged sum associated to the tree-level amplitudes reads
\begin{equation} \label{trsas}
	\begin{aligned}
		\left \langle d(n) \right \rangle \left ( \Tilde{\mathcal{Z}}_I \right )  &= \ell_I  \sum_{a\, |\, H_a <0} \ \text{Z}_I^a \ f_a(\ell_I,n)
		\\
		&=  \sqrt{\tfrac{\ell_I}{2}}  \sum_{a\, |\, H_a <0} \ \text{Z}_I^a  \, d_a(0)  \, \frac{\left | H_a \right |^{(d-1)/4}}{n^{(d+1)/4}} \, e^{ \frac{4\pi}{\ell_I} \, \sqrt{\left | H_a \right | \, n}} \ ,
	\end{aligned}
\end{equation}
and it grows exponentially if and only if a tachyonic character is present in the associated loop-channel amplitudes. If no tachyons are present in the tree-level (loop) channel, the sector averaged sum $\left \langle d(n) \right \rangle \left (\mathcal{Z}_I \right ) $ ($\left \langle d(n) \right \rangle  ( \Tilde{\mathcal{Z}}_I  ) $) vanishes identically. This result clearly extends the notion of misaligned supersymmetry to orientifold vacua.  

Notice that, albeit the vanishing of the sector averaged sum of the parent closed oriented string and of the $\left \langle d(n) \right \rangle  ( \Tilde{\mathcal{Z}}_I  ) $'s automatically guarantees the classical stability of the orientifold vacua, it is not true in general that a $\left \langle d(n) \right \rangle  ( \Tilde{\mathcal{Z}}_I  ) \neq 0$ implies that physical tachyons are present in the spectrum. In fact, tachyons could be projected away by $\Omega$, and when this happens each individual sector averaged sum can experience an exponential growth. Although one would naively expect that upon summing the contributions from all amplitudes ${\mathcal T}$ and $\Tilde{\mathcal Z}_I$ this exponential growth should disappear, this cannot be shown using the state of the art technology as will be extensively discussed in Section \ref{discussion}. Still, because the mere role of the Klein bottle and M\"obius strip amplitudes is to enforce the orientifold projection on ${\mathcal T}$ and ${\mathcal A}$, the simultaneous vanishing of the sector averaged sums associated to the torus and to the tree-level channel annulus amplitudes is a sufficient condition for the classical stability of the orientifold vacuum. 	
	
	\section{Ten-dimensional models} \label{10dmodels}
	
	The easiest and perhaps most interesting environment to test the previous results are the orientifold projections of the type IIB, 0A and 0B superstrings in $10$ dimensions. Indeed, these theories are quite simple to deal with, but nevertheless present all the subtleties we have discussed so far. The spectrum is organised according to the conjugacy classes of the ten-dimensional little group $SO(8)$, whose characters are dressed with the Dedekind $\eta$ functions from the non-compact bosons  \cite{AFL, Vafa}
	\begin{equation} \label{pseudoSO8}
		(O_{8} , V_{8} , S_{8} , C_{8} ) \to \left(\frac{O_{8} }{\eta^8}, \frac{V_{8}}{\eta^8} , \frac{S_{8}}{\eta^8} , \frac{C_{8}}{\eta^8}\right)\, .
	\end{equation}   
The torus partition function for the closed oriented strings can be written as 
	\begin{equation}
		\mathcal{T}= \sum_{a,b=1}^4 \mathcal{N}_{a b} \chi_a \bar{\chi}_b\ ,
	\end{equation}
and modular invariance, together with spin-statistics in space time, constrains the choice of the GSO matrix $\mathcal{N}_{ab}$. Aside from the type IIA superstring which will not play any role in what follows, the consistent choices for the GSO matrices are 
	\begin{equation} \label{clpart}
		\mathcal{N}^{(\text{IIB})}=\begin{pmatrix} 0 & 0 & 0 & 0 \\ 0 & 1 & -1 & 0 \\  0 & -1 & 1 & 0 \\ 0 & 0 & 0 & 0 \end{pmatrix} \ \ , \ \ \mathcal{N}^{(\text{0A})}= \begin{pmatrix} 1 & 0 & 0 & 0 \\ 0 & 1 & 0 & 0 \\  0 & 0 & 0 & 1 \\ 0 & 0 & 1 & 0 \end{pmatrix} \ \ \text{and} \ \ \mathcal{N}^{(\text{0B})}= \begin{pmatrix} 1 & 0 & 0 & 0 \\ 0 & 1 & 0 & 0 \\  0 & 0 & 1 & 0 \\ 0 & 0 & 0 & 1 \end{pmatrix}.
	\end{equation} 
These vacua are left-right symmetric on the world-sheet and thus can be modded out by the world-sheet parity $\Omega $ \cite{Sagnotti:1987tw, Pradisi:1988xd, Horava:1989vt, BS, Bianchi:1990tb, Bianchi:1991eu}, possibly combined with extra symmetries. 
Following the rules of the orientifold construction (see, for instance, \cite{AS}), the Klein bottle projection for the type IIB and type 0A superstrings is unique \cite{BS,Sagnotti1}, while four independent choices are allowed for the type 0B \cite{BS,Sagnotti1, Sagnotti2}, which provides an interesting arena to test our results. For this reason, we shall discuss in detail the behaviour of the sector averaged sum of the type 0B orientifolds, while we shall only briefly comment upon the type IIB and 0A cases. 
	
	\subsection{Type 0B orientifolds}
	
	As anticipated, the type 0B superstring is characterised by a diagonal GSO matrix and admits four ways to build a consistent orientifold projection \cite{BS,Sagnotti1,Sagnotti2}. These four possibilities find their justification on the fact that the choice of $\Omega$ is not unique, and actually can accommodate extra signs in the Klein bottle amplitude, compatibly with unitarity constraints \cite{Pradisi:1995qy, Pradisi:1995pp, Pradisi:1996yd}. One finds  
	\begin{equation} \label{klein0Bd}
		\mathcal{K}_A=  \sum_{a=1}^4 \ (-1)^{\sigma_A} 2 S_{A a} \ \chi_a \ , \qquad \sigma_A=\begin{cases}
			0 \ ,&  A=1,2
			\\
			1 \ ,&  A=3,4
		\end{cases} \ ,
	\end{equation}
	depending only on the form of the $S$ modular transformation \footnote{The $S$ modular transformation on the $SO(8)$ characters is $S= \frac{1}{2} \begin{pmatrix}
		1 & 1 & 1 & 1 \\
		1 & 1 & -1 & -1 \\
		1 & -1 & 1 & -1 \\
		1 & -1 & -1 & 1
	\end{pmatrix} $. For later use the $P$ transformation is $P=\text{diag}(-1,1,1,1)$.}, with $A=1,2,3,4$ denoting the $SO(8)$ conjugacy classes\footnote{ The amplitudes for $A=3,4$ are actually related by a parity operation, thus yielding equivalent results \cite{Sagnotti1,Sagnotti2}.} as ordered in \eqref{pseudoSO8}.  For the purpose of our analysis, it is important to notice that the classical stability of the closed string sector depends on the choice of $A$. Indeed, for $A=1,2$ the tachyon survives the orientifold projection, while for $A=3,4$ it is projected out. For these models the $S$ modular transformation yields 
		\begin{equation} \label{klein0Btr}
		\Tilde{\mathcal{K}}_A= 2^6 \ (-1)^{\sigma_A} \chi_A \ .
	\end{equation}
    This means, however, that only for the $A=1$ orientifold the tachyon couples to the O-planes, while it decouples in the $A=2,3,4$ cases, since only the dilaton or RR fields are present in $\Tilde{\mathcal K}_{2,3,4}$. 
   
The open-string sector is encoded in the annulus and M\"obius strip amplitudes, and depends on the choice of $\Omega$. One finds \cite{BS,Sagnotti1,Sagnotti2}
    \begin{equation} \label{an0Bd}
    	\mathcal{A}_A=   \sum_{a,b,c} \mathcal{N}_{a b}^c n_A^a n_A^b \ (-1)^{\sigma_{\left [c \right ] \times \left [ A \right ]}}  \chi_{\left [c \right ] \times \left [ A \right ]}  \ ,
    \end{equation}
 for the annulus amplitude, and 
       \begin{equation} \label{moebius0B}
   	\mathcal{M}_A=  \varepsilon_A \sum_a 2 S_{A a} n^a (-1)^{\sigma_a} \hat{\chi}_A \ ,
   \end{equation} 
for the M\"obius strip amplitude. In these expressions, $n_a$ denote the Chan-Paton labels, while $ \mathcal{N}_{a b}^c$ are the fusion rule coefficients, given by the Verlinde formula \cite{Verlinde:1988sn}
    \begin{equation} \label{verlinde}
    	\mathcal{N}_{a b}^c= \sum_m \frac{S_b^m S_a^m S_m^c }{ S_1^m }\ ,
    \end{equation} 
    reported in table \ref{fusionrules} for $SO(8)$. Finally, $\chi_{\left [c \right ] \times \left [ A \right ]}$ denotes the character obtained by the fusion $[c]\times [A]$.
    \begin{table}[ht!] \centering
    	\begin{tabular}{ |p{0.5cm}||p{2cm}|  } 
    		\hline
    		\  & $O_8 \ \  V_8 \ \ S_8 \ \ C_8 $ \\
    		\hline
    		\hline
    		$O_8$  & $O_8 \ \  V_8 \ \ S_8 \ \ C_8 $   \\
    		$V_8$  & $V_8 \ \  O_8 \ \ C_8 \ \ S_8 $  \\
    		$S_8$  & $S_8 \ \  C_8 \ \ O_8 \ \ V_8 $ \\
    		$C_8$  & $C_8 \ \  S_8 \ \ V_8 \ \ O_8 $  \\
    		\hline 
    	\end{tabular}
    	\caption{The fusion rules for $SO(8)$ characters.}
    	\label{fusionrules}
    \end{table}

    The tadpole conditions, which can be extracted from the tree-level amplitudes $\Tilde{\mathcal K}_A$ and
    \begin{equation} \label{anm0Btr}
    	\begin{aligned}
    		&\Tilde{ \mathcal{A}}_A= 2^{-6} \sum_a (-1)^{\sigma_{A}} 2 S_{A a} \bigg ( \sum_b 2 S_{a b} n_A^b (-1)^{\sigma_b} \bigg )^2  \chi_{a} \ ,
    		\\
    		& \Tilde{ \mathcal{M}}_A= 2 (-1)^{\delta_{A,1}}  \varepsilon_A \sum_b 2S_{A b} n_A^b (-1)^{\sigma_b} \ \hat{\chi}_A \ ,
    	\end{aligned}
    \end{equation}
    fix the signs\footnote{Strictly speaking, it is not mandatory to impose the inhomogeneous tadpole conditions for the $A=1,2$ models, since they are associated to NS-NS characters. However, in what follows we shall always impose their cancellation.} $\varepsilon_A=-(-1)^{\sigma_A}$ for all models, and determine the Chan-Paton gauge group. As a result, the gauge group for the model $A=1$ is $ G_1=U(n) \times U(n+32)$ with real tachyons in the $(A+\bar A, 1)+ (1,S+\bar S)$ representation, left-handed ($S_8$) fermions in the $(F, F ) + (\bar F , \bar F)$ representation, and right-handed fermions ($C_8$) in the  $(F, \bar F ) + (\bar F , F)$ representation. Here and in the following we denote by $F$ ($\bar F$) the (conjugate) fundamental representation and by $S$ ($A$) the (anti-)symmetric representations. 
    
For the model $A=2$, the gauge group is a product of orthogonal ones $G_2=[SO (n) \times SO(32-n)]^2$ and the massless spectrum comprises tachyons in the $(F,1,F,1)+(1,F,1,F)$ representation, together with left-handed and right-handed fermions in the $(F,1,1,F)+(1,F,F,1)$ and $(F,F,1,1)+(1,1,F,F)$ representations, respectively. 

Finally, the models $A=3$ and $A=4$ are equivalent up to a change in the chirality of the fermions, and yield a gauge group $G_{3,4} = U(32+n) \times U(n)$ with tachyons in the $(F, \bar F)+(\bar F , F)$ representation, fermions of one chirality in the $(A+\bar A,1)+(1,S+\bar S)$ representation and fermions of opposite chirality in the $(F,F)+(\bar F , \bar F)$. The choice $n=0$ removes the tachyon also from the open-string sector, which makes the type $0'\text{B}$ vacuum classically stable \cite{Sagnotti1,Sagnotti2}.
 
    Once the amplitudes are settled, we can proceed to compute the sector averaged sums. In light of our general discussion, the sector averaged sum associated to \eqref{klein0Bd} should be non vanishing only for $A=1$, since this is the only case in which the tachyonic character is present in the tree-level channel. Indeed, only the universal leading term contributes to \eqref{leadPhiKA} and thus $\langle d(n) \rangle \left (\mathcal{K}_A \right )$  depends on the combinations of signs characterizing the Klein bottle amplitude. This means that only for the $A=1$ case, where the signs are all positive, the sector averaged sum does not vanish,
	\begin{equation}
			\left \langle d(n) \right \rangle\left (\mathcal{K}_1 \right )= 2 \frac{e^{4 \pi \sqrt{n/2}}}{ (2n)^{11/4}} \ ,
	\end{equation} 
while for all the other cases
	\begin{equation}
			\left \langle d(n) \right \rangle\left (\mathcal{K}_A \right )=0 \ , \qquad \text{for $A=2,3,4$} \ ,
	\end{equation}
    since the number of $+1$'s and $-1$'s is the same. As stressed in the previous Section, this result simply shows the possibility for the closed-string tachyon to couple to orientifold planes. Therefore, whenever $\langle d (n) \rangle (\mathcal{K})\neq 0$ {\itshape necessarily} a closed string tachyon is part of the spectrum, and  cannot be projected out by $\Omega$. On the contrary, a vanishing sector averaged sum for ${\mathcal K}$ is not a sufficient condition for stability, as can be seen from model 2, where a tachyon survives the orientifold projection even though it does not couple to O-planes. We will come back on these points in Section \ref{discussion}. 
	
	The action of the world-sheet parity $\Omega$ on the closed-string tachyon is, instead, encoded in $\langle d(n) \rangle \left (\Tilde{\mathcal{K}}_A\right )$. Since the tachyon is always present in $\mathcal{K}_A$, we expect that $\left \langle d(n) \right \rangle \left ( \Tilde{\mathcal{K}}_A \right ) $ should always be non trivial. A direct computation shows
	\begin{equation} \label{0Btrklein}
		\begin{array}{ll}
			\left \langle d(n) \right \rangle\left (\Tilde{\mathcal{K}}_A \right )=\frac{e^{4 \pi \sqrt{n/2}}}{ (2n)^{11/4}}  & \text{for $A=1,2$} \ ,
			\\ 
			\\
			\left \langle d(n) \right \rangle\left (\Tilde{\mathcal{K}}_A \right )=-\frac{e^{4 \pi \sqrt{n/2}}}{  (2n)^{11/4}} & \text{for $A=3,4$} \ , 
		\end{array}
	\end{equation}
	which indeed receive the leading exponential contribution from $O_8$. From \eqref{0Btrklein} it is possible to see that the sector averaged sums encode the IR divergence of the direct amplitude, and the overall sign carry information about the action of $\Omega$ on $O_8$. However, although in models $A=3,4$ the closed tachyon is project away from the physical spectrum, $\left \langle d(n) \right \rangle\left (\Tilde{\mathcal{K}}_{3,4} \right )$ fails to cancel the contribution from the torus amplitude \cite{AFL} 
	\begin{equation} \label{tor0B}
		\langle d(n) \rangle \left ( \mathcal{T} \right )= \sum_{\ell=1}^{\infty} \varphi (\ell ) \, \frac{e^{\frac{8 \pi}{\ell} \sqrt{ n/2}}}{(2 n)^{11/2}}\,  ,
	\end{equation} 
and therefore it seems that the vanishing of the overall $\langle d (n) \rangle$ is not a necessary condition for classical stability in orientifold vacua. 	

	This result is not peculiar of the type 0B superstring and we shall further comment on it in Section \ref{discussion}.
	
	 A similar analysis can be performed for the open sector. The sector averaged sums associated to $\mathcal{A}_A$ encodes the coupling of closed tachyons to D-branes. A straightforward computation yields  
	 \begin{equation} \label{sasan0Bd}
	 	\begin{array}{ll}
	 		\left \langle d(n) \right \rangle \left (\mathcal{A}_A \right )=\frac{1}{2}\frac{e^{4 \pi \sqrt{n/2}}}{(2n)^{11/4}} (m+ \overline{m}-n - \overline{n})^2 & \text{for $A=1$} \ ,
	 		\\
	 		\\
	 		\left \langle d(n) \right \rangle \left (\mathcal{A}_A \right )=\frac{1}{2}\frac{e^{4 \pi \sqrt{n/2}}}{(2n)^{11/4}} (n_1 + n_2-n_3-n_4)^2  &  \text{for $A=2$} \ ,
	 		\\ 
	 		\\
	 		\left \langle d(n) \right \rangle \left (\mathcal{A}_A \right )= 0 &  \text{for $A=3,4$} \ ,
	 	\end{array}
	 \end{equation}
	 where $m$, $\bar m$, $n$ , $\bar n$ and $n_i$ label the Chan-Paton multiplicities associated to the various gauge groups, as in \cite{BS,Sagnotti1,Sagnotti2}.
The vanishing of $\left \langle d(n) \right \rangle \left (\mathcal{A}_{3,4} \right )$ reflects the (trivial) cancellation of the tachyon tadpole in $\tilde{\mathcal A}_{3,4}$.

Similarly, the coupling of the closed string tachyon to D-branes and O-planes is encoded in the sector averaged sums associated to the M\"obius strip amplitudes. Since only in the $A=1$ case the tachyon couples to cross-caps and therefore only in this case is present in $\tilde{\mathcal M}$, the sector averaged sums $	\left \langle d(n) \right \rangle\left (\mathcal{M}_{2,3,4} \right )$ vanish independently of the brane configuration. For $A=1$, instead,
	 \begin{equation}\label{sasmoebius0Bd1}
	 		\left \langle d(n) \right \rangle\left (\mathcal{M}_1 \right )= \frac{e^{4 \pi \sqrt{n/2}}}{2^{9/4} \  n^{11/4}} (n + \overline{n}-m - \overline{m}) \ ,
	 \end{equation}
which can never vanish, unless the tadpole associated to $O_8$ is relaxed and $n=m$. 	 

Notice that the sector averaged sums depend on the choice of the Chan-Paton multiplicities via the tadpole of $O_8$. Unless this is uniquely fixed by the tadpole conditions, as in the $A=3,4$ models, one has the freedom to properly choose the Chan-Paton multiplicities so that the sector averaged sum can be set to zero. This is the case, for instance, of eqs. \eqref{sasan0Bd} and \eqref{sasmoebius0Bd1}. However, this does not yield any information on the presence or absence of closed string tachyons, since  $\langle d(n) \rangle \left ( \mathcal{Z}_I \right ) =0$ is a necessary but not a sufficient condition for classical stability. 

We now move to the sector averaged sums associated to $\Tilde{ \mathcal{A}_A}$ and $\Tilde{ \mathcal{M}_A}$, which encode the presence of open-string tachyons in the spectrum and their orientifold projection, respectively. 

A direct computation
   \begin{equation} \label{0Btrannulus}
   	\begin{array}{ll}
   		\left \langle d(n) \right \rangle \left (\Tilde{\mathcal{A}}_1 \right )=\frac{e^{4 \pi \sqrt{n/2}}}{  (2 n)^{11/4}} (n^2+\overline{n}^2 + m^2 + \overline{m}^2) \ ,
   		\\
   		\\
   		\left \langle d(n) \right \rangle \left (\Tilde{\mathcal{A}}_2 \right )=\frac{e^{4 \pi \sqrt{n/2}}}{  (2n)^{11/4}} 2(n_1 n_2 + n_3 n_4) \ ,
   		\\ 
   		\\
   		\left \langle d(n) \right \rangle \left (\Tilde{\mathcal{A}}_A \right )= \frac{e^{4 \pi \sqrt{n/2}}}{  (2n)^{11/4}} 2 (n_A^v \overline{n_A^o} +  \overline{n_A^v} \ n_A^o ) \ , & \text{for $A=3,4$} \ ,
   	\end{array}
   \end{equation}
clearly shows that only the $\ell=1$ terms in the Rademacher expansion does contribute to $\Tilde{ \mathcal{A}_A}$, consistently with the properties of the Kloosterman sums discussed in the previous Section. The dependence of \eqref{0Btrannulus} on the Chan-Paton factors reproduces the coefficients of $O_8$ in the annulus amplitudes. 

In the case of $\Tilde{ \mathcal{M}_A}$, 
   \begin{equation} \label{0Btrmoebius}
   	\begin{array}{ll}
   		\left \langle d(n) \right \rangle\left (\Tilde{\mathcal{M}}_1 \right )=-\frac{e^{2 \pi \sqrt{n/2}}}{ 2^{9/4} n^{11/4}} (n + \overline{n} - m-\overline{m}) \ ,
   		\\ 
   		\\
   		\left \langle d(n) \right \rangle\left (\Tilde{\mathcal{M}}_A \right )=0 \ , & \text{for $A=2,3,4$} \ ,
   	\end{array}
   \end{equation}
only the $\ell =2$ term contributes, which involves the $P$ transformation connecting the two description of the M\"obius strip. This is the reason why only $\left \langle d(n) \right \rangle\left (\Tilde{\mathcal{M}}_1 \right )$ is non-vanishing, since in the $A=2,3,4$ vacua the tachyon transforms in bi-fundamental representations and therefore is oriented. 
The choice $n_{3,4}^v =0$ is compatible with the tadpole conditions and yields the non-tachyonic type $0'\text{B}$ vacuum of \cite{Sagnotti1, Sagnotti2}. In this case, the vanishing of $\left \langle d(n) \right \rangle \left (\Tilde{\mathcal{A}}_{3,4} \right )$ is a sufficient condition for classical stability. 

Open-string tachyons can also be absent in model 1, if the NS-NS tadpoles associate to $O_8$ is relaxed. In this case, one can choose $m=0$ and $n=1$, yielding a $U(1)$ gauge group. The open-string tachyons would still transform in the antisymmetric representations which, however, is zero-dimensional for $U(1)$. Therefore, it is the orientifold projection, {\em i.e.} an interplay between ${\mathcal A}_1$ and ${\mathcal M}_1$, which removes the open-string tachyon. However, $\langle d (n) \rangle (\Tilde{\mathcal A}_1 )$ and $\langle d (n) \rangle (\Tilde{\mathcal M}_1 )$ fail to cancel each other which, again, shows that it is impossible to compare sector averaged sums associated to different Riemann surfaces, as discussed in Section \ref{discussion}.

Our results are compatible with \cite{Cribiori2} for the type  $0'\text{B}$ superstring, although from our analysis and from that of \cite{Niarchos1,Niarchos2}, valid for open strings, the vanishing of $\langle d(n) \rangle \left ( \mathcal{Z}_I \right )$ is interpreted as the decoupling of closed tachyons to D-branes and O-planes. 
	
	\subsection{Comments on type 0A and type IIB orientifolds}

	A similar study can be easily performed in the case of orientifold projections of the type 0A, the type I superstring and the Sugimoto vacuum with $n_\pm$ stacks of D-branes and $\overline{D}$-branes \cite{BS,Sugimoto:1999tx}. Following the analysis of the previous section, one finds that the sector averaged sums for the loop amplitudes of the type I superstring and Sugimoto model are always vanishing, consistently with \cite{Cribiori2}, since tachyons are not present in the closed-string spectrum, and therefore cannot couple to O$_\pm$-planes and (anti)D-branes. Moving to the sector averages sums associated to the tree-level amplitudes, the only one to be non-zero is  
	\begin{equation}
		\left \langle d(n) \right \rangle\left (\Tilde{\mathcal{A}} \right )=\frac{e^{4 \pi \sqrt{n/2}}}{  (2n)^{11/4}} 2 \, n_+ \, n_- \ ,
	\end{equation} 
reflecting the instability of the brane-antibrane system, which includes an open-string tachyon in the bi-fundamental representation. 

The orientifold of the type 0A superstring \cite{BS} does not reveal any further insight. Indeed, the tachyonic character is always present in every amplitude and, following the conventions in \cite{BS} for the Chan-Paton labels, the corresponding sector averaged sums read 
	\begin{equation}
		\begin{array}{ll}
			\left \langle d_{0A}(n) \right \rangle\left (\mathcal{K} \right )= \frac{e^{4 \pi \sqrt{n/2}}}{  (2 n)^{11/4}} \ , & \left \langle d_{0A}(n) \right \rangle\left (\Tilde{\mathcal{K}} \right )=\frac{e^{4 \pi \sqrt{n/2}}}{ (2 n)^{11/4}} \ ,
			\\ 
			\\
			\left \langle d_{0A}(n) \right \rangle\left (\mathcal{A} \right )= \frac{e^{4 \pi \sqrt{n/2}}}{(2n)^{11/4}} \left ( n_B -n_F \right )^2 \ , & \left \langle d_{0A}(n) \right \rangle\left (\Tilde{\mathcal{A}} \right )= \frac{e^{4 \pi \sqrt{n/2}}}{(2 n)^{11/4}} \left ( n_B^2 + n_F^2 \right ) \ ,
			\\
			\\
			\left \langle d_{0A}(n) \right \rangle\left (\mathcal{M} \right )= - \frac{e^{2 \pi \sqrt{n/2}}}{  2^{9/4} n^{11/4}} \left ( n_B -n_F \right ) \ , & \left \langle d_{0A}(n) \right \rangle\left (\Tilde{\mathcal{M}} \right )=\frac{e^{2 \pi \sqrt{n/2}}}{ 2^{9/4} n^{11/4}} \left ( n_B -n_F \right ).
		\end{array}
	\end{equation}
	
	\section{Scherk-Schwarz orientifolds} \label{Scherk-Schwarz orientifolds}
	Another interesting class of non-supersymmetric vacua can be constructed in lower dimensions by compactifying the type IIB superstring {\em \`a la} Scherk-Schwarz \cite{Scherk:1979zr, Ferrara:1987es,Kounnas:1989dk,Scherk:1978ta} on a circle $S^1 (R)$ of radius $R$. This model can be realised as a freely acting $\mathbb{Z}_2$ orbifold whose generator is  $g= (-1)^F \delta$, where $F$ is the space-time fermion number and $\delta$ shifts the compact coordinate $y$ as $  y \to y +\pi R$ . This orbifold interpolates between the type 0B vacuum as $R\to 0$ and the supersymmetric type IIB theory as $R \to \infty$. This construction preserves the left-right symmetry of the theory and therefore admits an orientifold construction, as was done in \cite{Antoniadis:1998ki}. 
In order to study the sector averaged sum using the tools introduced in the previous section, it is convenient to take rational values for the compactification radius \cite{AFL}
	\begin{equation} \label{ratrad}
		\frac{R^2}{\alpha'}=\frac{s}{t} \ ,\qquad \text{with}\quad \text{gcd}(s,t)=1\ ,
	\end{equation} 
so that the contribution of the Narain lattice collapses to a finite number of characters, and the CFT becomes rational. To avoid subtleties with the proper definition of the characters on the freely acting orbifold \cite{AFL}, we shall restrict our discussion to even values of $s$, and without loss generality we can also take\footnote{The more general case is presented in the Appendix \ref{rational Scherk-Schwarz}} $t=1$. The associated characters read
	\begin{equation} \label{lattchar}
		\lambda_a (q) = \sum_m \frac{q^{s \left( m + \frac{a}{2s }\right)^2}}{\eta (q)} \ \qquad a=0,\ldots , N-1\,,
	\end{equation}
and have (shifted) conformal weights $H_a = \alpha^2/4s -1/24$. Among the $N=2s$ characters, $\lambda_0$ and $\lambda_{s}$ are real, while $\lambda_a$ and $\lambda_{2s-a}$, $a=1,\ldots , s-1$,  form conjugate pairs. 

The full RCFT is built upon the tensor product of the $SO(8)$ and the lattice characters
	\begin{equation} \label{pseudochar}
		\{\chi_\alpha \}_{\alpha=0}^{8s-1} = (O_8 , V_8 ,  S_8 ,  C_8 ) \otimes \{\lambda_a \}_{a=0}^{2s-1}\,,
	\end{equation}
which have conformal weights $H_{2sp+a} = \frac{a^2}{4s} - \frac{1}{2} \delta_{p,0}$, with $p=0,\ldots ,3$. Notice that $H_a <0$ for $a < \sqrt{2s}$ and $p=0$, so that the associated characters can describe tachyonic states.

The torus amplitude for this freely acting orbifold reads \cite{AFL}
	\begin{equation} \label{torSS}
		\begin{split}
			{\mathscr Z} &= \sum_{a=0}^{s-1}\left(  |\chi_{2a+ 2s}|^2 + |\chi_{2a+4s}|^2 \right) 
			\\
			&\quad - \sum_{a=0}^{s-2} \left( \chi_{2a+1+2s} \bar \chi_{2a+1+4s} + \chi_{2a+1+4s}\, \bar\chi_{2a+1+2s} \right)
			\\
			&\quad +\sum_{a=0}^{\frac{s}{2}-1} \left( \chi_{2a+\sigma}\, \bar\chi_{2a+\sigma+s} + \chi_{2a+\sigma+s}\, \bar\chi _{2a+\sigma} \right.
			\\
			&\qquad\qquad\qquad \left.
			+ \chi_{2a+\sigma+6s}\, \bar\chi_{2a+\sigma+7s} + \chi_{2a+\sigma+7s}\, \bar\chi _{2a+\sigma+6s} \right)
			\\
			&\quad -\sum_{a=0}^{\frac{s}{2}-1} \left( \chi_{2a+1-\sigma} \, \bar \chi_{2a+1-\sigma+7s} + \chi_{2a+1-\sigma+7s}\, \bar \chi_{2a+1-\sigma} \right.
			\\
			&\qquad \qquad \qquad\left.
			+\chi_{2a+1-\sigma+s} \, \bar\chi_{2a+1-\sigma+6s} + \chi_{2a+1-\sigma+6s} \, \bar \chi_{2a+1-\sigma+s}
			\right) \,,
		\end{split}
	\end{equation}
where in the third and fourth sums the two cases $s=2(2m+\sigma)$ with $\sigma=0,1$ have to be considered separately. The study of misaligned supersymmetry for this oriented closed string vacuum was already performed in \cite{AFL}, and we shall not repeat it here. Instead, we shall analyse the properties of its orientifold projections, starting from the canonical Klein bottle amplitude 
	\begin{equation} \label{kleinSS}
		\begin{split}
			{\mathcal K} &=  \sum_{a=0}^{s-1}\left(  \chi_{2a+ 2s} \ - \ \chi_{2a+4s} \right)  \, ,
		\end{split}
	\end{equation}
that symmetrises the NS-NS states from $|\chi_{2a+ 2s}|^2$ and anti-symmetrises the R-R ones from $|\chi_{2a+4s}|^2 $. Upon an $S$ modular transformation, one gets the tree-level amplitude
	\begin{equation} \label{trkleinSS}
		\begin{split}
			\Tilde{\mathcal K} &= 2^4 \sqrt{s}
			\sum_{b=0}^{1}\left(  \chi_{bs+ 2s} \ - \ \chi_{bs+4s} \right)\, .
		\end{split}
	\end{equation}
The characters $\chi_{2s} = V_8\lambda_0$ and $\chi_{4s} = S_8 \lambda_0$ are massless and therefore induce non-trivial tadpoles which can be cancelled introducing suitable configurations of D-branes. To parallel the discussion in \cite{Antoniadis:1998ki}, we introduce two stacks $n_1,n_2$ of D-branes and two stacks $n_3,n_4$ of $\overline{\text{D}}$-branes, so that 
	\begin{equation}
		\begin{aligned}
			\Tilde{\mathcal{A}}= 2^{-6}\sqrt{s} \  \sum_{b=0}^{1} & \bigg \{  \left ( n_1 + n_2 + n_3 + n_4 \right )^2 \chi_{2s+bs}  - \left ( n_1 + n_2 - n_3 - n_4 \right )^2    \chi_{4s+bs}   \\
			& + \left ( n_1 - n_2 + n_3 - n_4 \right )^2 \chi_{\tfrac{2 b+1}{2}s}   - \left ( n_1 - n_2 - n_3 + n_4 \right )^2 \chi_{6s+ \tfrac{2 b+1}{2}s} \bigg \}.
		\end{aligned}
	\end{equation} 
and
	\begin{equation} \label{moebiusSS}
			\Tilde{\mathcal{M}}= - \sqrt{s} \  \sum_{b=0}^{1}  \bigg \{  \left ( n_1 + n_2 + n_3 + n_4 \right ) \hat{\chi}_{2s+bs}  - \left ( n_1 + n_2 - n_3 - n_4 \right )  (-1)^b  \hat{\chi}_{4s+bs}   \bigg \} \,,
	\end{equation} 
which is written in terms of the real hatted characters \cite{BS,AS}. The tadpole conditions for the massless states are
\begin{equation} \label{SStadpole}
\begin{aligned}
&  n_1 + n_2 + n_3 + n_4 = 32\,, & \qquad \text{from} \qquad \chi_{2s}\ ,
\\
&  n_1 + n_2 - n_3 - n_4 = 32\,, & \qquad \text{from} \qquad \chi_{4s}\ ,
\end{aligned}
\end{equation}
as in \cite{Antoniadis:1998ki}. The signs of the Chan-Paton factors in the $\chi_{4s}$ tadpole clearly reveals that branes and anti-branes have opposite RR charge, while from the first tadpole we see that they all have the same tension. 

The loop-channel annulus and M\"obius strip amplitudes can be obtained via an $S$ and $P$ modular transformation, respectively, and read
	\begin{equation}
		\begin{aligned}
			\mathcal{A}= &  \left ( n_1^2+n_2^2+n_3^2+n_4^2 \right )  \sum_{a=0}^{s/2-1} \left [   \chi_{2s+4 a } -  \chi_{4s + 4a+2}\right ]  \\
			& + 2 \left ( n_1 n_2 + n_3 n_4 \right ) \sum_{a=0}^{s/2-1} \left [   \chi_{2s+2 +4a } -   \chi_{4s + 4a}\right ] \\
			& + 2 \left ( n_1 n_3 + n_2 n_4 \right ) \sum_{a=0}^{s/2-1} \left [  \chi_{4 a } - \chi_{6s + 2+4a}\right ] \\
			& + 2 \left ( n_1 n_4 + n_2 n_3 \right ) \sum_{a=0}^{s/2-1} \left [  \chi_{2+4 a } -  \chi_{6s + 4a}\right ],
		\end{aligned}
	\end{equation}
and
	\begin{equation} \label{trmoebiusSS}
			\mathcal{M} = -   \sum_{a=0}^{s/2-1} \left \{ \left ( n_1 + n_2 + n_3 + n_4 \right )\hat{\chi}_{4 a+2s}  - \left ( n_1 + n_2 - n_3 - n_4 \right )   \hat{\chi}_{ 4a+2+4s} \right \}.
	\end{equation}
The light spectrum of this orientifold vacuum includes a graviton, a RR two-form and the dilaton from the closed-string sector together with vectors with gauge group $SO(n_1)\times SO(n_2) \times SO(n_3) \times SO(n_4)$, right-handed fermions in the representation $(n_1, 1 ; 1 , n_4 )+ (1,n_2 ;  n_3 ,1 )$ and tachyons in the representation $(n_1, 1 ;  n_3  ,1)+ (1,n_2 ;1,  n_4 )$, from the open-string sector. Additional tachyonic states are present for special values of the compactification radius. In the open sector they emerge when $R^2/\alpha '  = s > 2$ and they come in the representation $(n_1, 1 ; 1 , n_4 )+ (1,n_2 ;  n_3 ,1 )$, while in the closed sector they are present for $R^2 /\alpha ' = s < 8$. 

As shown in Appendix \ref{rational Scherk-Schwarz}, these amplitudes are clearly compatible with those of \cite{Antoniadis:1998ki} once the value $R^2 = s\alpha '$ is chosen, and the momentum and winding sums are written in terms of the $\lambda$ characters. 

We can now proceed to study the sector averaged sums associated to this vacuum. We start from the behaviour of the massive degrees of freedom in the loop channels, which reflect the presence/absence of closed-string tachyons freely propagating between D-branes and orientifold planes. From the explicit expressions \eqref{trkleinSS} and \eqref{moebiusSS} we see that the transverse-channel Klein bottle and M\"obius strip amplitudes only involve massless or massive states, and therefore the associated sector averaged sums, $\langle d (n) \rangle ({\mathcal K})$ and $\langle d (n) \rangle ({\mathcal M})$, vanish identically, as can be seen from a direct computation. 
Different is the case of the annulus amplitude. In fact, the sector averaged sum $\langle d (n) \rangle ({\mathcal A})$ vanishes only for $s \geq 8$, whereas for $s < 8$ is given by
	\begin{equation}
		\left \langle d(n) \right \rangle \left ( \mathcal{A} \right )= 2 \times\frac{ \sqrt{s}}{2\sqrt{2}} \frac{1}{ n^{\frac{5}{2}}} \left ( \frac{8-s}{16} \right )^2 e^{\pi \sqrt{(8-s)\, n} } \left (n_1-n_2+n_3-n_4 \right )^2.
	\end{equation}
	This result is consistent with the data present in the transverse amplitude, since the only {\itshape would be} tachyonic characters that can propagate between boundaries are $\chi_{s/2}$ and $\chi_{3s/2}$, and for $s\geq 8$, they are massive. Notice that the sector average sum depends on the tachyonic tadpole and thus vanishes if $n_1-n_2+n_3-n_4 =0$. 
	
	 Turning to the classical stability of open strings, this is related to the large-mass behaviour of the associated tree-level amplitudes. The transverse channel M\"obius  has a vanishing sector averaged sum, as expected from the fact the open-string tachyons transform into bifundamental representations of $SO(n_1)\times SO(n_2) \times SO(n_3) \times SO(n_4)$, and thus do not contribute to ${\mathcal M}$. The sector averaged sum for the transverse annulus is 
	 \begin{equation} \label{trsasannSS1}
	 	\begin{aligned}
	 		\left \langle d(n) \right \rangle \left ( \Tilde{\mathcal{A}} \right )= \frac{1}{\sqrt{2} n^{\frac{5}{2}}} &\left \{ 2 \left ( n_1 n_3 + n_2 n_4 \right ) \sum_{a < \sqrt{\frac{s}{8}}} \left ( \frac{8 a^2 -s}{2s} \right )^2   \ e^{4 \pi \sqrt{\left |8a^2 - s \right | \, n/2s}}\right. \\
	 		&\left. + 2 \left ( n_1 n_4 + n_2 n_3 \right )  \sum_{a < \sqrt{\frac{s}{8}}-\frac{1}{2}} \left ( \frac{2 (2a+1)^2-s}{2s} \right )^2   \ 
			e^{4 \pi \sqrt{\left |2 (2a+1)^2-s \right | \, n/2s}} \right \} \  ,
	 	\end{aligned}
	 \end{equation}
and reveals the presence of open tachyons stretched between branes and anti-branes \footnote{The range of $a$ in the sums in \eqref{trsasannSS1} is a short-hand notation to indicate the sum over $a$ associated to characters with $H_{4a}<0$ and $H_{4a+2}<0$ respectively. }. Although for arbitrary values of the Chan-Paton multiplicities the first contribution is always present, the second disappears whenever $s\leq 2$ since in this case the Wilson line relative between the $n_1$ ($n_2$) branes and the $n_4$ ($n_3$) anti-branes makes the tachyon massive. As a result, there are two interesting solutions of the tadpole conditions \eqref{SStadpole} which make $\langle d(n) \rangle (\tilde{\mathcal A})=0$:  $n_1 +n_2  = 32$, $ n_3 = n_4 =0$ where only D-branes are introduced, and $n_1 - n_4 =32$, $n_2=n_3=0$,  where the branes and anti-branes have a relative Wilson line, which becomes tachyonic when $s\leq 2$. In these cases, the vanishing of the sector averaged sums of the transverse annulus is a necessary and sufficient condition for the classical stability of the open-string sector. This is so, because the would-be tachyons are oriented and therefore $\langle d (n) \rangle (\tilde{\mathcal M} )\equiv 0$, always.
 
Finally, it is easy to show that $\left \langle d(n) \right \rangle \left ( \Tilde{\mathcal{K}} \right )=0$, which implies that the closed-string tachyons cannot be eliminated via the $\Omega$ projection. As a result, also in this case, the necessary and sufficient condition for the classical stability of the closed-string sector is the vanishing of 
$\left \langle d(n) \right \rangle \left ( {\mathcal{T}} \right )$ which occurs when $s\ge 8$.

	\section{M-theory orientifolds} \label{M-theory orientifolds}
			
	The Scherk-Schwarz compactification studied in the previous Section is a natural generalisation of what one naturally does in field theory \cite{Scherk:1978ta, Scherk:1979zr}. In the language of freely-acting orbifolds \cite{Kounnas:1989dk} it combines the action of the space-time fermion number with a shift $\delta:\  y \to y+\pi R$ of the compact direction. 
	
String Theory, however, allows for a different construction where $(-1)^F$ is accompanied by the asymmetric shift $\tilde\delta: \ y_{L,R} \to y_{L,R} \pm \frac{\alpha'}{2 R} \pi$, which now affects the winding modes. This action has a natural interpretation in terms of the T-dual variable $\tilde y = y_L-y_R$. Clearly, this variation of Scherk-Schwarz breaking does not bring anything new when studying closed oriented strings. However, new physics emerge in the orientifold case, since T-duality changes the dimensionality of the orientifold planes and D-branes, and also the $\Omega$ projection. This new scenario was called {\itshape M-theory breaking}  \cite{Antoniadis:1998ki,Antoniadis:1998ep} and we will study its misaligned supersymmetry. To do this, we have to restrict to rational values of the compactification radius
	\begin{equation}
		R^2=\frac{s}{t}\alpha' \ ,
	\end{equation} 
with $s,t$ co-prime.

Similarly to the Scherk-Schwarz breaking case, the compatibility between the shift $\tilde\delta: \ y_{L,R} \to y_{L,R} \pm \frac{\alpha'}{2 R} \pi$ and the character decomposition of the Narain lattice is non-trivial: for $t$ even the $N=2st$ characters $\lambda_a$ in \eqref{lattchar} are eigenstates of $\tilde\delta$ and thus suffice for the construction of the vacuum. For $t$ odd, instead, a refinement in terms of $4N$ characters is needed, as described in Appendix \ref{rational Scherk-Schwarz}. For simplicity, in the following we shall take $t$ even and $s=1$. The general case will be presented in Appendix \ref{rational M-theory}. 

The torus amplitude associated to the $(-1)^F \, \tilde \delta$ orbifold of the type IIB superstring reads
	\begin{equation} \label{torM}
		\begin{split}
			{\mathscr Z} &= \sum_{a=0}^1 \left ( \left | \chi_{2t+at} \right |^2 + \left | \chi_{4t+at} \right |^2 \right )
			\\
			&+ \sum_{b=1}^{\frac{t}{2}-1}\left(  \chi_{2t+2b} \bar{\chi}_{2t-2b}+ \chi_{4t+2b}\bar{\chi}_{4t-2b} \right.
			\\
			&\qquad\qquad\qquad \left.
			+ \chi_{2t+t+2b}\, \bar\chi_{2t-t-2b} + \chi_{4t+t+2b}\, \bar\chi _{4t-t-2b} \right)
			\\
			&-\sum_{b=0}^{\frac{t}{2}-1}\left(  \chi_{2t+2b+1} \bar{\chi}_{4t-2b-1}+ \chi_{4t+2b+1}\bar{\chi}_{2t-2b-1} \right.
			\\
			&\qquad\qquad\qquad \left.
			+ \chi_{2t+t+2b+1}\, \bar\chi_{4t-t-2b-1} + \chi_{4t+t+2b+1}\, \bar\chi _{2t-t-2b-1} \right)
			\\
			&+  \sum_{a=0}^{1} \left ( \left | \chi_{\frac{2a+1}{2} t} \right |^2 + \left | \chi_{6t+\frac{2a+1}{2}t}\right |^2 \right )
			\\
		    &+ \sum_{b=1}^{\frac{t}{2}-1}\left(  \chi_{\frac{t}{2}-2b} \bar{\chi}_{\frac{t}{2}+2b}+ \chi_{6t+\frac{t}{2}-2b} \bar{\chi}_{6t+\frac{t}{2}+2b}\right.
		    \\
		    &\qquad\qquad\qquad \left.
		    + \chi_{\frac{3t}{2}-2b}\, \bar\chi_{-\frac{t}{2}+2b} + \chi_{6t+\frac{3t}{2}-2b}\, \bar\chi_{6t-\frac{t}{2}+2b} \right)
		    \\
		     &- \sum_{b=0}^{\frac{t}{2}-1}\left(  \chi_{\frac{t}{2}-2b-1} \bar{\chi}_{\frac{t}{2}+2b+1}+ \chi_{6t+\frac{t}{2}-2b-1} \bar{\chi}_{6t+\frac{t}{2}+2b+1}\right.
		     \\
		     &\qquad\qquad\qquad \left.
		     + \chi_{\frac{3t}{2}-2b-1}\, \bar\chi_{-\frac{t}{2}+2b+1} + \chi_{6t+\frac{3t}{2}-2b-1}\, \bar\chi_{6t-\frac{t}{2}+2b+1} \right) \, .
		\end{split}
	\end{equation}
This construction preserves world-sheet parity and thus  can be orientifolded by adding the Klein bottle amplitude that, for the standard choice of $\Omega$, reads
	\begin{equation} \label{kleinM}
		\begin{split}
			{\mathscr K} &=  \sum_{a=0}^{1}\left(  \chi_{a t+ 2t} - \chi_{at+4t} \right)
			+ \sum_{a=0}^{1} \left( \chi_{\frac{2a+1}{2} t}\, - \chi_{\frac{2a+1}{2}t+ 6 t}\right ) \,.
		\end{split}
	\end{equation}
Notice that aside from the first two terms associated to the $V_8$ and $S_8$ characters dressed with the KK and winding excitations, two extra contributions appear associated to $O_8$ and $C_8$.  In the transverse channel
	\begin{equation} \label{trklein}
		\Tilde{\mathcal{K}}=  2^5 \frac{2}{\sqrt{t}} \sum_{b=0}^{t/2-1} \left \{ \chi_{2t+4b}- \chi_{4t+4b+2} \right \} \ ,
	\end{equation}
so that this vacuum involves pairs of $O9_-$ and $\overline{O9}_-$ planes which do not carry a net RR charge, as can be seen from the fact that the characters $\chi_{4t+4b+2}= S_8 \, \lambda_{4b+2}$ are massive, $m^2\propto \frac{(4b+2)^2}{4t} = \frac{1}{4} R^2\, (4 b+2)^2 $. In the formal limit $R\to 0$ they become massless, the $\overline{O9}_-$ planes decouple and a net RR tadpole emerges from \eqref{trklein}. Taking into account this limiting case,  we shall also impose the RR tadpoles associated to the $\chi_{4t+4b+2}$ characters.

Following \cite{Antoniadis:1998ki}, we add $(n_1,n_2)$ stacks of branes and $(n_3,n_4)$ stacks of antibranes so that
	\begin{equation}\label{trannM}
		\begin{aligned}
			\Tilde{\mathcal{A}}= 2^{-5} \frac{2}{\sqrt{t}} \sum_{b=0}^{t/2-1} & \left \{  \left ( n_1 + n_2 + n_3 + n_4 \right )^2  \chi_{2t + 4 b}  + \left ( n_1 - n_2 + n_3 - n_4 \right )^2 \chi_{2t+ 2 + 4b }  \right.  
			\\
			& \left.  -  \left ( n_1 + n_2 - n_3 - n_4 \right )^2 \chi_{4t +4b}- \left ( n_1 - n_2 - n_3 + n_4 \right )^2  \chi_{4t+4b+2}\right \} \ ,  
		\end{aligned}
	\end{equation}
and
	\begin{equation} 
		\begin{aligned}
			\Tilde{\mathcal{M}}= - 2 \frac{2}{\sqrt{t}} \sum_{b=0}^{t/2-1} & \left \{  \left ( n_1 + n_2 + n_3 + n_4 \right )  \hat{\chi}_{2t + 4 b}   -   \left ( n_1 - n_2 - n_3 + n_4 \right ) \hat{\chi}_{4t+4b+2}\right \}.   
		\end{aligned}
	\end{equation}  

The tadpole conditions 
\begin{equation}
n_1 + n_2 + n_3 + n_4 = 32\ , \qquad n_1 + n_2 - n_3 - n_4 = 0\ ,
\end{equation}
associated to the massless characters $\chi_{2t}$ and $\chi_{4t}$, together with the ``massive'' tadpole $n_1 - n_2 - n_3 + n_4 = 32$ for $\chi_{4t+4b+2}$, admit the unique solution
\begin{equation}
n_1 = n_4 =16\ , \qquad n_2 = n_3 =0\,.
\end{equation}
$S$ and $P$ modular transformations yield the loop amplitudes
	\begin{equation} \label{annM}
		\begin{aligned}
			\mathcal{A}= \sum_{a=0}^{1} \bigg \{   & \left ( n_1^2+n_2^2 + n_3^2+n_4^2\right )  \left (   \chi_{2t + \frac{2 a }{2} t}  -\chi_{4t + \frac{2 a }{2} t}\right )     
			 \\
			& + 2 \left ( n_1 n_2 + n_3 n_4 \right )  \left (   \chi_{2t + \frac{2 a +1}{2} t}  -\chi_{4t + \frac{2 a +1 }{2} t}  \right )     
			 \\
			 & + 2 \left ( n_1 n_3 + n_2 n_4 \right )  \left (   \chi_{ \frac{2 a }{2} t}  -\chi_{6t + \frac{2 a  }{2} t}  \right )     
			 \\
			& +  2 \left ( n_2 n_3 + n_1 n_4 \right ) \ \left (  \chi_{\frac{2 a+1 }{2} t}  - \chi_{6t+\frac{2 a+1 }{2} t} \right ) \bigg \} \ ,
		\end{aligned}
	\end{equation}
	and 
	\begin{equation}
		\begin{aligned}
			\mathcal{M}= -  \sum_{a=0}^{1}& \left \{ \left ( n_1 + n_2 + n_3 + n_4\right )  \hat{\chi}_{2t+\frac{2 a }{2} t}  -  \left ( n_1 - n_2 - n_3 + n_4\right ) (-1)^a \hat{\chi}_{4t+\frac{2 a }{2} t}  \right \} \ .
		\end{aligned}
	\end{equation}
	
At low energy, this M-theory breaking orientifold describes a graviton, the dilaton and a RR two-form from the closed string sector together with an $SO(16) \times SO(16)$ gauge group coupled to a left-handed fermion in the adjoint representation. For large values of the compactification radius, {\em i.e.} for $t<8$ a real closed-string tachyon and an open-string one in the $(16,16)$ representation emerge. 

As shown in Appendix \ref{rational M-theory}, these amplitudes are clearly compatible with those of \cite{Antoniadis:1998ki} once the value $R^2 = \alpha '/t$ is chosen, and the momentum and winding sums are written in terms of the $\lambda$ characters. 
	
	Given this M-theory breaking vacuum, we can study its classical stability by computing the associated sector averaged sums. An immediate inspection of the direct channel amplitudes gives $\left \langle d(n) \right \rangle \left ( \mathcal{Z}_I \right )=0$, reflecting the absence of closed string tachyons that freely propagate  between D-branes and/or O-planes. This, however, does not say anything about the classical stability of the whole construction, since a tachyon can be present in ${\mathcal T}$ for special values of $t$. A complete different story regards the sector averaged sum associated to the transverse channel. 
	In fact, from $\Tilde{\mathcal K}$ one finds
	\begin{equation}
	\left \langle d(n) \right \rangle \left ( \Tilde{\mathcal{K}} \right )= 2 \frac{1}{\sqrt{2} n^{\frac{5}{2}}} \left ( \frac{8-t}{16} \right )^2 e^{\pi \sqrt{(8-t)\, n}}\ ,
\end{equation}
for $t<8$, while it vanishes if $t \geq 8$. This behaviour is consistent with the fact that the closed-string spectrum is tachyon-free for $t \geq 8$, while a tachyon is present when $t<8$.

The sector averaged sum from the transverse annulus amplitude has a similar behaviour
\begin{equation} \label{sastrannM}
	\begin{aligned}
		\left \langle d(n) \right \rangle \left ( \Tilde{\mathcal{A}} \right )=  2 \frac{1}{\sqrt{2} n^{\frac{5}{2}}}  \ 2 n_1 n_4 \  \left ( \frac{8-t}{16}  \right )^2 e^{ \pi \sqrt{(8-t)\, n} }  ,
	\end{aligned}
\end{equation}
reflecting the presence of a tachyon in the bi-fundamental $(n_1,n_4)$ representation when $t<8$. When $t \geq 8$, $\left \langle d(n) \right \rangle \left ( \Tilde{\mathcal{A}} \right )=0$ since all would-be tachyons are actually massive. Finally, the sector averaged sum of the transverse Moebius strip amplitude vanishes identically, consistently with the fact that the open-string tachyon, when present, is oriented transforming in the bifundamental representations of $SO(16) \times SO(16)$. Therefore, as in the case of the Scherk-Schwarz breaking, the vanishing of the sector averaged sum $\langle d (n) \rangle (\tilde{\mathcal A})$ is a necessary and sufficient condition for the classical stability of the open-string spectrum, while for the stability of the closed-string sector we must require that both $\langle d (n)\rangle ({\mathcal T})$ and  $\langle d (n)\rangle (\tilde{\mathcal K})$ vanish. 

\subsection{A variation on M-theory breaking orientifolds} \label{varMtheory}

The Klein bottle of eq. \eqref{kleinM} is not the only orientifold projection compatible with \eqref{torM}. Indeed, in \cite{Dudas:2002dg, Dudas:2000sn, Dudas:2003wp} it was shown that other choices of $\Omega$ are possible, and in particular the amplitude 
\begin{equation} \label{kleinM'}
	{\mathscr K}' =  \sum_{a=0}^{1}\left(  \chi_{a t+ 2t} - \chi_{at+4t} \right)
	- \sum_{a=0}^{1} \left( \chi_{\frac{2a+1}{2} t}\, - \chi_{\frac{2a+1}{2}t+ 6 t}\right ) \, ,
\end{equation}
has the virtue of projecting away the closed string tachyon. This change of sign affects the geometry of the orientifold planes, since now
\begin{equation} \label{trkleinM'}
	\Tilde{\mathcal{K}}'=  2^5 \frac{2}{\sqrt{t}} \sum_{b=0}^{t-1} \left \{ \chi_{2t+4b+2}- \chi_{4t+4b} \right \} \ ,
\end{equation}
which reveals the presence of $O9_-$ and $\overline{O9}_+$ planes which have a net RR charge but vanishing tension. The annulus amplitude is still given by eqs. \eqref{annM} and \eqref{trannM}, while the tree-level channel M\"obius strip amplitude reads
\begin{equation} \label{trmoebiusM'} 
	\begin{aligned}
		\Tilde{\mathcal{M}}'= - 2 \frac{2}{\sqrt{t}} \sum_{b=0}^{t-1} & \left \{  \left ( n_1 - n_2 + n_3 - n_4 \right )  \hat{\chi}_{2t + 4 b+2}   -   \left ( n_1 + n_2 - n_3 - n_4 \right ) \hat{\chi}_{4t+4b}\right \}.   
	\end{aligned}
\end{equation}  
The RR tadpole conditions
\begin{equation}
	n_1 + n_2 - n_3 - n_4 =32 \ , \qquad n_1 - n_2 - n_3 + n_4 =0 
\end{equation}
 are clearly incompatible with the NS-NS tadpole $n_1 + n_2 + n_3 + n_4 =0$ since the O-planes are tensionless. Upon a $P$ modular transformation, we get  
\begin{equation} \label{moebiusM'}
	\begin{aligned}
		\mathcal{M}'= -  \sum_{a=0}^{1}& \left \{ \left ( n_1 - n_2 + n_3 - n_4 \right ) (-1)^a \hat{\chi}_{2t+\frac{2 a }{2} t} - \left ( n_1 + n_2 - n_3 - n_4 \right ) \hat{\chi}_{4t+\frac{2 a }{2} t}  \right \} \ .
	\end{aligned}
\end{equation}    

The light spectrum from the closed strings comprises a graviton, the dilaton and a RR two-form, and no extra light states emerge at special values of the compactification radius, since the would-be tachyon is removed by the new orientifold projection. From \eqref{annM} and \eqref{moebiusM'} we read the gauge group $SO(n_1) \times USp(n_2) \times SO(n_3) \times USp(n_4)$. Left-handed fermions and tachyons transform in the representations\footnote{$S$ ($A$) denotes the (anti-)symmetric representation of the orthogonal/symplectic gauge group.} $(A_1,1;1,1)+(1,A_2;1,1)+(1,1;S_3,1)+(1,1;1,S_4)$ and $(n_1,1;n_3,1) + (1,n_2;1,n_4)$, respectively. When $t<8$ additional tachyons in the representations $(n_1,1;1,n_4) + (1,n_2;n_3,1) $ appear. Clearly, all open-string ta\-chyons can be eliminated by taking the minimal solution of the RR tadpoles, $n_1=n_2=16$ and $n_3=n_4=0$, but we prefer to consider the more general set-up with branes and anti-branes to show a richer behaviour of the sector average sum. 
It is possible to show that, choosing the value $R^2 = \alpha '/t$ in the model of \cite{Dudas:2002dg} one obtains the amplitudes presented in this section.

As in the case of the standard M-theory breaking, $\langle d(n) \rangle (\mathcal{Z}'_I) \equiv 0$ for every loop amplitude, since no closed-string tachyon propagates between D-branes and/or O-planes. The sector averaged sums associated to the tree-level channels are, instead, slightly modified and read 
\begin{equation} \label{sastrkleinM'}
	\left \langle d(n) \right \rangle \left ( \Tilde{\mathcal{K} }' \right )= - 2 \frac{1}{\sqrt{2} n^{\frac{5}{2}}} \left ( \frac{8-t}{16} \right )^2 e^{\pi \sqrt{(8-t)\, n} } \ ,
\end{equation}
and 
\begin{equation} \label{sastrannM'}
		\left \langle d(n) \right \rangle \left ( \Tilde{\mathcal{A}}' \right )=   \frac{1}{\sqrt{2} n^{\frac{5}{2}}}  \left \{ 2\left (  n_1 n_4 + n_2 n_3 \right )   2  \left ( \frac{8-t}{16} \right )^2 e^{\pi \sqrt{ (8-t) \, n}}  + 2\left (  n_1 n_3 + n_2 n_4 \right ) \frac{1}{4}\,  e^{4 \pi \sqrt{n/2}} \right \},
\end{equation}
while $\left \langle d(n) \right \rangle \left ( \Tilde{\mathcal{M}}' \right )=  0 $ since, also in this case, open-string tachyons are oriented. Despite the only difference with the standard M-theory breaking analysis is in the minus sign in eq. \eqref{sastrkleinM'}, the physical interpretation is {\em deeply} different. It is telling us that the tachyon that appears in the spectrum when $t < 8$ is actually projected away. However, eq. \eqref{sastrkleinM'} fails to cancel the sector averaged sum of the torus
\begin{equation} \label{sastorMth}
	\langle d(n) \rangle ( \mathcal{T})= \left (\frac{8-t}{16} \right )^4 \frac{1}{ n^{5}}\, 
	\sum_{\ell=1}^{\infty} \, \varphi (\ell) \, e^{\frac{2\pi}{\ell} \sqrt{(8-t)\, n}} \ ,
\end{equation}
because of the appearance of the sub-leading contributions, $\ell >1$. It is tempting to say that the minus sign of $\left \langle d(n) \right \rangle \left ( \Tilde{\mathcal{K}}' \right )$ is a necessary condition for the stability of the closed-string sector, even though the complete sector averaged sum does not vanish. 

To conclude,  eq. \eqref{sastrannM'} reflects the presence of tachyons in the spectrum, as long as anti-branes are present. Also in this case, since all tachyons transform in bi-fundamental representations, a necessary and sufficient condition for the stability of the open-string sector is the vanishing of  $\left \langle d(n) \right \rangle \left ( \Tilde{\mathcal{A}}' \right )$.

	\section{Discussion}\label{discussion}
	
	In Section \ref{Exact} we have shown in full generality that the sector averaged sum of a given amplitude grows exponentially whenever tachyonic characters are present in the corresponding dual amplitude, obtained by inverting the proper time. As a result, $\langle d(n) \rangle (\mathcal{Z}_I)$ carries information about the coupling of closed string tachyons to D-branes and/or orientifold planes, while $\langle d(n) \rangle \left ( \Tilde{\mathcal{Z}}_I \right ) $ encodes the presence of tachyonic characters in the one loop partition functions. It is then clear that it is the latter which is closely related to the classical stability of the vacuum. Still, contrary to what happens in oriented closed strings \cite{AFL}, the vanishing of each $\langle d(n) \rangle \left ( \Tilde{\mathcal{Z}}_I \right ) $ is not a necessary and sufficient condition for its stability. Actually, the conditions $\langle d(n) \rangle ( \mathcal{T}) =0 $ and $\langle d(n) \rangle ( \Tilde{\mathcal{A}}) =0$ are sufficient but {\em not} necessary. This is so because the Klein bottle and M\"obius strip amplitudes enforce the $\Omega$ projection, and thus if tachyons are absent in ${\mathcal T} $ and ${\mathcal A}$ they are automatically absent in ${\mathcal K}$ and ${\mathcal M}$. On the contrary, no information can be drawn from the vanishing of $\langle d(n) \rangle ( \Tilde{\mathcal{K}}) $ and $\langle d(n) \rangle ( \Tilde{\mathcal{M}})$, since tachyons could be associated to oriented strings. 
Moreover, if $\langle d(n) \rangle ( \mathcal{T}) \neq 0 $ and {\em all} $\langle d(n) \rangle \left ( \Tilde{\mathcal{Z}}_I \right ) \neq 0$ one cannot conclude that the vacuum is unstable, since tachyons could be projected away by the orientifold projection. This means, that they are present in each amplitude, and appear in ${\mathcal K}$ and ${\mathcal M}$ with a minus sign. In this case, one could be tempted to conclude that the conditions for classical stability be $\langle d(n) \rangle ( \mathcal{T})  + \langle d(n) \rangle ( \Tilde{\mathcal{K}}) =0$ and $\langle d(n) \rangle ( \Tilde{\mathcal{A}}) +\langle d(n) \rangle ( \Tilde{\mathcal{M}}) =0$. However, this could not be the case since one cannot achieve complete cancellation between different amplitudes. In fact, let us look at the sector averaged sum \eqref{largemasssas} associated to the torus amplitude. It contains infinite contributions with $\ell \geq 1$, which clearly cannot be cancelled by $\langle d(n) \rangle ( \Tilde{\mathcal{K}})  $, which instead has only the leading term, $\ell=1$. A simple example of this kind is the orientifold projection with $A=3$ or $4$ in Section  \ref{10dmodels}. In these cases, the closed-string tachyon is projected away by $\Omega$, but 
	\begin{equation} \label{0Bavsumtor}
		 \langle d(n) \rangle \left ( \mathcal{T} \right ) = \sum_{\ell=1}^\infty \varphi(\ell) \frac{e^{\frac{8 \pi}{\ell}\sqrt{n/2}}}{(2n)^{11/2}} \ ,
	\end{equation} 
and
	 \begin{equation} \label{0Bavsumkl}
	 		\left \langle d(n) \right \rangle\left (\Tilde{\mathcal{K}}_{3,4} \right )=-\frac{e^{4 \pi \sqrt{n/2}}}{ (2n)^{11/4}} \ ,
	 \end{equation} 
and thus can never cancel each other.	Another example showing the same feature is provided by the alternative orientifold construction for the $9d$ M-theory breaking vacua described in Section \ref{varMtheory}, in which  \eqref{sastorMth} can never be cancelled by \eqref{sastrkleinM'}.

	A similar story goes on for the open string sector. Because of the properties \eqref{Kloostann} and \eqref{Kloostmob} of the generalised Kloosterman sums, only one term contributes to the sector averaged sum, and it has $\ell=1$ for the transverse annulus but $\ell =2$ for the transverse M\"obius strip amplitude, and thus they can never cancel. As an example, we can consider the open sector of model $A=1$ in Section  \ref{10dmodels}. Tadpole conditions are compatible with $n=\bar n =1$ and $m=\bar m =0$, a choice that eliminates the open-string tachyon, since it now transforms in the antisymmetric representation of an $U(1)$ gauge group. In this case,
		\begin{equation} 
			 \left \langle d(n) \right \rangle \left (\Tilde{\mathcal{A}}_1 \right )=  \frac{e^{4 \pi \sqrt{n/2}}}{ (2 n)^{11/4}} \, 2 ,
			\end{equation}
		and
		\begin{equation}
			\left \langle d(n) \right \rangle\left (\Tilde{\mathcal{M}}_1 \right )= - \frac{e^{2 \pi \sqrt{n/2}}}{ 2^{9/4} \  n^{11/4}} \, 2 ,
	\end{equation} 
which clearly do not cancel each other. 	
	
	This suggests that the sector averaged sums associated to different Riemann surfaces cannot be directly compared. This is not surprising since they compute different properties of the vacuum. On the one hand, $\langle d(n) \rangle ({\mathcal T})$ controls the asymptotic growth of physical degrees of freedom and therefore it is tied to the spectrum of oriented closed strings. On the other hand, the quantities $\langle d(n) \rangle (\Tilde{\mathcal Z}_I)$ carry no global information on the spectrum, but rather to the one-point couplings of closed-string states to D-branes and/or O-planes. 	
	
In conclusion, these considerations show that the sector averaged sums have nothing to say for vacua in which classical stability is a consequence of a non trivial action of the orientifold projection.

	\section{Conclusions} \label{conclusion}
   
   In this paper we have addressed the issue of classical stability, {\em i.e.} absence of physical tachyons, in orientifold vacua. To this end, we have employed and further developed the tools of Misaligned Supersymmetry introduced in \cite{AFL, Cribiori1,Cribiori2,Dienes:1994np}.  The main result can be summarised as follows: the exponential growth of the number of degrees of freedom of the one loop partition functions is related to infrared divergences of the tree-level amplitudes, and {\em vice versa}. Per se this is not surprising since this UV/IR interplay descends from the dual interpretation of the Klein bottle, annulus and M\"obius strip amplitudes as loop or tree-level diagrams. 
However, this relation allows us to interpret the vanishing of the sector averaged sum associated to the loop channel as the decoupling of closed-string tachyons from D-branes, as first noted by \cite{Niarchos1, Niarchos2}, and from O-planes, thus providing a {\em necessary}, but not sufficient, condition for its absence from the physical spectrum. This result is consistent with the ten-dimensional examples of \cite{Cribiori2} and explicitly shows the connection of the sector averaged sum for the loop channel amplitude with the free propagation of tachyons between D-branes and O-planes, revealing the true physical meaning of $\langle d(n) \rangle$. 

Similarly, the sector averaged sum associated to the tree-level amplitudes is connected to the classical stability of the vacuum since it carries information about the presence of tachyons in the one-loop partition functions. Therefore, the vanishing of the sector averaged sums associated to both the torus and transverse annulus amplitudes is a {\em sufficient}, but not necessary, condition for the absence of tachyons, since the role of the Klein bottle and M\"obius strip amplitudes is to implement the orientifold projection, so that $\langle d (n) \rangle (\Tilde{\mathcal K})$ and $\langle d (n) \rangle (\Tilde{\mathcal M})$ must vanish whenever $\langle d (n) \rangle ({\mathcal T})=0 $ and $\langle d (n) \rangle (\Tilde{\mathcal A})=0$.

The obstruction in making Misaligned Supersymmetry a {\em necessary and sufficient} condition for classical stability resides in the fact that in orientifold vacua tachyons, although present in all amplitudes, can be removed by the orientifold projection. In these cases, Misaligned Supersymmetry would require non-trivial cancellations among amplitudes with different topologies. However, the sector averaged sums for the torus and the tree-level $\Tilde{\mathcal K}$, $\Tilde{\mathcal A}$ and $\Tilde{\mathcal M}$ amplitudes are built on CFT's on different Riemann surfaces and thus carry different information. As a result, it is difficult, if not impossible, to compare them. 

This means that {\em Misaligned Supersymmetry}, as we understand it, is not the right tool to uncover the reason behind the IR finiteness of the vacuum energy for general orientifold vacua, leaving such question still open. New ideas and technologies need to be developed to fully address this issue.

\section*{Acknowledgements}
The author is deeply grateful to Carlo Angelantonj for the intensive discussions about the role played by misaligned supersymmetry in non-supersymmetric orientifold constructions, as well as for his precious comments and feedbacks on the manu\-script. It is also a pleasure to thank Ivano Basile for helpful comments on the manuscript and Keith Dienes, Niccol\`o Cribiori, Flavio Tonioni, Susha Parameswaran and Timm Wrase for useful comments on the first version of the paper. The author would like to thank the Department of Theoretical Physics at CERN, the Max Planck Institute of Munich and the Centre of Theoretical Physics at the \'Ecole Polytechnique for hospitality during various stages of this project.

	\begin{appendix}
		
		\section{Rational Scherk-Schwarz partition functions}
		\label{rational Scherk-Schwarz}
		As discussed in \cite{AFL}, for rational values of the radius, $R^2=\frac{s}{t} \alpha'$, the Narain lattice for a circle compactification collapses to a RCFT involving a finite number $N=2st$ of characters,
		\begin{equation}\label{lattch}
			\lambda_{a}= \frac{1}{\eta} \sum_{k \in \mathbb{Z}} q^{\frac{N}{2}( k + \frac{a}{N})^2} \ ,
		\end{equation}
which, under the generators of the modular group, transform with the matrices
		\begin{equation}
				T_{a b}= e^{i \pi \left (\tfrac{a^2}{2st}-\tfrac{1}{12} \right )} \delta_{a b} \ ,\qquad 	S_{a b}= \frac{e^{ \tfrac{2 \pi i a b}{N}}}{\sqrt{N}}\ . 
		\end{equation} 
In general, the $\lambda$'s are not compatible with the free action of the Scherk-Schwarz mechanisms. In fact, as shown in \cite{AFL}, for $s$ odd  one needs to introduce the refined characters
		\begin{equation}\label{xichar}
			\xi_a=  \frac{1}{\eta(\tau)} \sum_{k} q^{2 N \left ( k + \frac{a}{4 N} \right )^2} \ ,
		\end{equation}
where now $a=0, \ldots , 4N-1$, and the modular matrices acting upon them read
		\begin{equation}
				T_{a b}= e^{i \pi \left (\tfrac{a^2}{8st}-\tfrac{1}{12} \right )} \delta_{a b} \ , \qquad 	S_{a b}= \frac{e^{ \tfrac{2 \pi i a b}{4N}}}{\sqrt{4N}} \ . 
		\end{equation}

Following \cite{Vafa}, the Narain lattice partition function reads
\begin{equation}
\sum_{m,n} \Lambda_{m,n}= \sum_{a=0}^{N-1} \lambda_a \bar{\lambda}_{l a} = \sum_{\sigma,\rho=0}^1 \sum_{a=0}^{N-1} \xi_{2(a + \sigma N)} \bar{\xi}_{2(l a + \rho N)}\ ,
\end{equation}
with $l=rt+vs $, and $r,v$ solutions of $rt-vs=1$. The action of the half-shift on these characters, and on the twisted sector with half-integer windings reads
		\begin{equation}\label{SSlatticeeven}
			\begin{aligned}
				&\sum_{m,n} (-1)^m \Lambda_{m,n}=\sum_{a=0}^{N-1} (-1)^{\frac{1+l}{2t}a} \lambda_{a} \bar{\lambda}_{l a},
				\\
				&	\sum_{m,n}\Lambda_{m, n + \frac{1}{2}} = \sum_{ r=0}^{N-1}  \lambda_{\Tilde{r}} \bar{\lambda}_{r}, 
				\\
				&\sum_{m,n}(-1)^m \Lambda_{m, n + \frac{1}{2}} = \sum_{r=0}^{N-1} (-1)^{\frac{\Tilde{r}^2 - r^2}{N}} \lambda_{\Tilde{r}} \bar{\lambda}_{r},
			\end{aligned}
		\end{equation}
		for even $s$, where $\Tilde{r}= l r -\tfrac{1+l}{2}s$,  and  
		\begin{equation}\label{SSlatticeodd}
			\begin{aligned}
				&\sum_{m,n} (-1)^m \Lambda_{m,n}=\sum_{\sigma,\rho=0}^1 \sum_{a=0}^{N-1} (-1)^{\sigma+\rho + \tfrac{1+l}{2t} a} \ \xi_{2(a + \sigma N)} \bar{\xi}_{2(l a + \rho N)},
				\\
				&	\sum_{m,n}\Lambda_{m, n + \frac{1}{2}} = \sum_{r, c=0}^{2N-1} 
				\xi_{2r+1} \bar{\xi}_{2 c + 1} \ \delta_{r,\hat{c}}^{(N)}, 
				\\
				&\sum_{m,n}(-1)^m \Lambda_{m, n + \frac{1}{2}} = \sum_{c, r=0}^{2N-1} (-1)^{\frac{(2r+1)^2 - (2 c+1)^2}{4 N}}
				\xi_{2r+1} \bar{\xi}_{2 c + 1} \ \delta_{r,\hat{c}}^{(N)},
			\end{aligned}
		\end{equation}
		for odd $s$. In equation \eqref{SSlatticeeven} $\Tilde{r}= l r -\tfrac{1+l}{2}s$, while in eq. \eqref{SSlatticeodd} the Kronecker delta imposes $r=\hat{c}$ where $\hat{c}= l c -\tfrac{1+l}{2} s-\tfrac{1-l}{2} $ mod $N$.

		\subsection{Klein bottle amplitudes}
		\label{rational Scherk-Schwarz klein}
		 For an arbitrary radius the Klein-bottle amplitude encoding the orientifold projection on the closed strings reads \cite{Antoniadis:1998ki}
		\begin{equation}
			\mathcal{K}= \left ( V_8 - S_8 \right ) P_{2 m} (q^2)
		\end{equation}
		where
		\begin{equation}
			P_{2m}(q^2)= \frac{1}{\eta} \sum_{m\in\mathbb{Z}} q^{\frac{\alpha'}{2} \left (\frac{2m}{R} \right )^2}.
		\end{equation}
For rational values of the radius, one finds 
		\begin{equation}
			P_{2m}(q^2) = \frac{1}{\eta} \sum_{a=0}^{s-1}  \sum_{k\in \mathbb{Z}}\left ( q^2 \right )^{\frac{N}{2} \left (k + \frac{a}{s} \right )^2}  = \sum_{a=0}^{s-1} \lambda_{2 a t}
		\end{equation}
for even $s$, and 

		\begin{equation}
				P_{2m}(q^2)	= \frac{1}{\eta}  \sum_{a=0}^{s-1}  \sum_{k\in \mathbb{Z}} \left ( \left ( q^2 \right )^{2 N \left (k + \frac{2a}{2 s} \right )^2} + \left ( q^2 \right )^{2 N \left (k + \frac{2a + 1}{2 s} \right )^2} \right ) 
= \sum_{\sigma=0}^{1} \sum_{a=0}^{s-1} \xi_{4t(2a + \sigma )}.
		\end{equation}
for odd values of $s$.

In the transverse-channel amplitude
		\begin{equation}
			\Tilde{\mathcal{K}}= \frac{2^5}{2} \  \left ( V_8 - S_8 \right ) W_n(q)
		\end{equation}
one finds
		\begin{equation}
				W_n(q)  = \frac{1}{\eta} \sum_n q^{\frac{\alpha'}{4}\left (\frac{n R}{\alpha'}\right )^2}  = 
				\begin{cases} \sum_{b=0}^{2t-1} \lambda_{bs} & \text{for $s$ even}\ ,
				\\
				\sum_{\sigma =0}^1 \sum_{b=0}^{2t-1} \xi_{2(2b +  \sigma  ) s} & \text{for $s$ even}\ .\end{cases}
		\end{equation}

		\subsection{Annulus and M\"obius amplitudes}
		\label{rational Scherk-Schwarz annulus}
		
Moving to the open sector, for arbitrary value of the radius $R$  the Annulus and Moebius strip amplitudes read \cite{Antoniadis:1998ki}
		\begin{equation}
			\begin{aligned}
				\mathcal{A}= & \left ( n_1^2+n_2^2+n_3^2+n_4^2 \right ) \left [V_8 P_{2m}(\sqrt{q}) - S_8 P_{2m+1}(\sqrt{q}) \right ] \\
				& + 2 \left ( n_1 n_2 + n_3 n_4 \right ) \left [ V_8 P_{2m+1}(\sqrt{q}) - S_8 P_{2m} (\sqrt{q}) \right ] \\
				& + 2 \left ( n_1 n_3 + n_2 n_4 \right ) \left [ O_8 P_{2m}(\sqrt{q}) - C_8 P_{2m+1} (\sqrt{q}) \right ] \\
				& + 2 \left ( n_1 n_4 + n_2 n_4 \right ) \left [ O_8 P_{2m+1}(\sqrt{q}) - C_8 P_{2m} (\sqrt{q}) \right ] 
			\end{aligned}
		\end{equation}
		and 
		\begin{equation}
			\mathcal{M}= -  \left \{ \left ( n_1 + n_2 + n_3 + n_4 \right ) \hat{V}_8 P_{2m}(\sqrt{q})  - \left ( n_1 + n_2 - n_3 - n_4 \right ) \hat{S}_8 P_{2m+1}(\sqrt{q}) \right \} \ .
		\end{equation}
For rational values of the radius,
		\begin{equation}
		(-1)^{m \, \beta} \, P_{m}(\sqrt{q}) = \frac{1}{\eta} \sum_{m} (-1)^{m\beta}\, q^{\frac{\alpha'}{2} \left ( \frac{m}{R} \right )^2} = 
		\begin{cases}		
		\sum_{a=0}^{s-1} (-1)^{a\beta}\, \lambda_{2 a t} & \text{for $s$ even}\ ,
		\\
		\sum_{\sigma=0}^1 \sum_{a=0}^{s-1} (-1)^{\sigma\beta} \xi_{4t(2a + \sigma )} & \text{for $s$ odd}\ ,
		\end{cases}
		\end{equation}
so that
	\begin{equation}
	P_{2m+\delta}  (\sqrt{q}) = \begin{cases} \sum_{a=0}^{s/2-1}  \lambda_{2t(2 a +\delta )} & \text{for}\ s\ \text{even}\ ,\\
				\sum_{a=0}^{s-1}  \xi_{4t(2a+\delta )} & \text{for}\ s\ \text{odd}\, ,\end{cases}
	\end{equation}
with $\delta =0, 1$. 
		
In the transverse channel, the amplitudes valid for generic values of the radius $R$ read
		\begin{equation}
			\begin{aligned}
				\Tilde{\mathcal{A}}= 2^{-5} \frac{v}{2} & \left \{ \left [ \left ( n_1 + n_2 + n_3 + n_4 \right )^2 V_8  
				 - \left ( n_1 + n_2 - n_3 - n_4 \right )^2 S_8 \right ]  W_n  \right.
				\\
				& \left. + \left [ \left ( n_1 - n_2 + n_3 - n_4 \right )^2 O_8  
				 - \left ( n_1 - n_2 - n_3 + n_4 \right )^2 C_8 \right ] W_{n + \frac{1}{2}} \right \} \ ,
			\end{aligned}
		\end{equation} 
and
		\begin{equation}
			\Tilde{\mathcal{M}}= - v \left \{ \left ( n_1 + n_2 + n_3 + n_4 \right ) \hat{V}_8 W_n  - \left ( n_1 + n_2 - n_3 - n_4 \right ) \hat{S}_8 (-1)^n W_n \right \}.
		\end{equation}
and call for the decompositions
\begin{equation}
W_{n+\frac{\delta}{2}} = \begin{cases} \sum_{b=0}^{2t-1} \lambda_{\frac{2 b+\delta}{2}s} &  \text{for $s$ even}\ ,
\\
\sum_{\sigma=0}^1 \sum_{b=0}^{2t-1} \xi_{(4 b  + 2 \sigma +\delta )s} &  \text{for $s$ odd}\ ,\end{cases}
\end{equation}
with $\delta =0,1$, once $R^2$ takes rational values. 

It is straightforward to see that these decompositions  reproduce the amplitudes used in Section \ref{Scherk-Schwarz orientifolds}, obtained from the associated torus amplitude via the implementation of the ``orientifold algorithm''.  
		
		\section{Rational M-theory breaking partition functions}
		\label{rational M-theory}
		This sections is devoted to the construction of the partition functions for the M-theory breaking when the compactification radius takes rational values, as done in Appendix \ref{rational Scherk-Schwarz}. Although the procedure follows similar steps, some care is needed since a different shift $\tilde\delta$ is now involved. This model is related by T-duality to the Scherk-Schwarz realization described previously, so that the results obtained in this context are T-dual to the previous ones. Since T-duality exchanges $s$ and $t$, two different cases arise according to the parity of $t$. For $t$ even, the characters $\lambda$ are eigenstates of $\tilde \delta$, 
	\begin{equation}
	\tilde\delta\cdot \lambda_a = (-1)^{a/2s}\, \lambda_a\ ,
	\end{equation}
and thus suffice to build the various amplitudes,
		\begin{equation}
			\begin{aligned}
				&(-1)^n \Lambda_{m,n}= \sum_{a=0}^{N-1} (-1)^{\frac{1- l}{2 s} a} \lambda_a \bar{\lambda}_{l a}\ , \\
				&\Lambda_{m + \frac{1}{2}, n }= \sum_{ r=0}^{N-1}  \lambda_{\Tilde{r}} \bar{\lambda}_{r} \ ,\\
				&(-1)^n \Lambda_{m+ \frac{1}{2}, n } = \sum_{r=0}^{N-1} (-1)^{\frac{\Tilde{r}^2 - r^2}{N}} \lambda_{\Tilde{r}} \bar{\lambda}_{r}\ ,
			\end{aligned}
		\end{equation}
where now $\Tilde{r}= l r -\frac{1-l}{2}t$. 

		For $t$ odd, the $\lambda$'s are no-longer eigenvalues of $\tilde\delta$ and must be split into the $4N$ characters $\xi_a$ of eq. \eqref{xichar}, so that
		\begin{equation}
			\begin{aligned}
				&(-1)^n \Lambda_{m,n}= \sum_{\sigma,\rho=0}^1 \sum_{a=0}^{N-1} (-1)^{\sigma+\rho + \frac{1-\lambda}{2s} a} \xi_{2(a + \sigma N)} \bar{\xi}_{2(\lambda a + \rho N)}\ ,\\
				&\Lambda_{m + \frac{1}{2}, n }=  \sum_{b,c=0}^{2N-1} \xi_{2b+1} \xi_{2c+1}\ \delta_{\hat{b},c}^{(N)},, \\
				&(-1)^n \Lambda_{m+ \frac{1}{2}, n } = \sum_{ b,c=0}^{2N-1} (-1)^{\frac{(2b+1)^2 - (2 c+1)^2}{4 N}}
				\xi_{2b+1} \bar{\xi}_{2 c + 1} \ \delta_{\hat{b},c}^{(N)}, 
			\end{aligned}
		\end{equation}
with $\hat{b}= l c - \frac{1-l}{2}t - \frac{1-l}{2}$. 
		
		\subsection{Klein bottle amplitudes}
		\label{rational M-theory klein}
		For an arbitrary radius, the Klein-bottle amplitude for the M-theory breaking reads \cite{Antoniadis:1998ki}
		\begin{equation}\label{kleinMR}
			\mathcal{K}= \left ( V_8 - S_8 \right ) P_m +  \left ( O_8 - C_8 \right ) P_{m+\frac{1}{2}} .
		\end{equation}
Taking $R^2 = \alpha' s/t$, a simple computation yields
\begin{equation}
P_{m+\frac{\delta}{2}}(q^2)  = \frac{1}{\eta} \sum_{m \in \mathbb{Z}} q^{\frac{t}{2s} \left (m+ \frac{\delta}{2} \right )^2}
				=\begin{cases} \sum_{a=0}^{2s-1} \lambda_{ (2a+ \delta) t/2} & \text{for $t$ even}\ ,
				\\ 
				\sum_{\sigma=0}^{1} \sum_{a=0}^{2s-1} \xi_{(4a+ \delta+ 2\sigma )t}& \text{for $t$ odd}\ .\end{cases}
\end{equation}
Similarly, in the transverse channel amplitude one finds
		\begin{equation}
			\Tilde{\mathcal{K}}= 2^5  2v \left \{ V_8 W_{4 n}(q) - S_8 W_{4 n + 2}(q) \right \}.
		\end{equation}
with, now, 
		\begin{equation}\label{wsumratMth}
				W_{4 n+2\delta} (q) = \frac{1}{\eta} \sum_{n\in\mathbb{Z}} q^{\frac{s}{t} \left (2n +\delta\right )^2} 
				= \begin{cases} \sum_{b=0}^{t/2-1} \lambda_{(4 b+ 2 \delta) s} & \text{for $t$ even}\ ,\\
				 \sum_{b=0}^{t-1 } \xi_{4( 2b +\delta) s} & \text{for $t$ odd}\ .
				 \end{cases}
		\end{equation}
		
		\subsection{Annulus and M\"obius amplitudes}
		\label{rational M-theory annulus}

		Moving to the open sector, for arbitrary values of the radius $R$ the annulus and M\"obius strip amplitudes read \cite{Antoniadis:1998ki}
		\begin{equation}
			\begin{aligned}
				\mathcal{A}= & \left [ \left ( n_1^2+n_2^2+n_3^2+n_4^2 \right )  \left ( P_{m}(\sqrt{q}) + P_{m+\frac{1}{2}}(\sqrt{q}) \right ) \right.  \\
				&\left.  + 2\left ( n_1 n_2 + n_3 n_4 \right ) \left ( P_{m + \frac{1}{4}}(\sqrt{q}) + P_{m+\frac{3}{4}}(\sqrt{q}) \right ) \right ] \left (V_8 - S_8 \right ) \\
				& + \left [ 2\left ( n_1 n_3 + n_2 n_4 \right ) \left ( P_{m}(\sqrt{q}) + P_{m+\frac{1}{2}}(\sqrt{q}) \right ) \right.   \\
				&\left.  + 2 \left ( n_1 n_4 + n_2 n_3 \right ) \left ( P_{m + \frac{1}{4}}(\sqrt{q}) + P_{m+\frac{3}{4}}(\sqrt{q}) \right ) \right ] \left (O_8 - C_8 \right ) \ ,
			\end{aligned}
		\end{equation}
and
		\begin{equation}
			\begin{aligned}
				\mathcal{M}= -  & \left \{ \left ( n_1 + n_2 + n_3 + n_4 \right ) \ \hat{V}_8 \left ( P_m(\sqrt{q}) + P_{m+\frac{1}{2}}(\sqrt{q}) \right ) \right. \\
				& \left. - \left ( n_1 - n_2 - n_3 + n_4 \right ) \hat{S}_8 \left ( P_{m}(\sqrt{q}) - P_{m+\frac{1}{2}}(\sqrt{q}) \right ) \right \}
			\end{aligned}
		\end{equation}
For rational values of the radius,
		\begin{equation}
				P_{m + \frac{\delta}{4}}(\sqrt{q}) = \frac{1}{\eta} \sum_{m} q^{\frac{t}{2s} \left (m + \frac{\delta}{4} \right )^2} \\
				=\begin{cases} \sum_{a=0}^{s-1} \lambda_{\frac{4 a + \delta}{2} t} & \text{for $t$ even}\ ,
			\\
				\sum_{\sigma=0}^1 \sum_{a=0}^{s-1} \xi_{t(8 a + \delta + 4 \sigma )} & \text{for $t$ odd}\ .
				\end{cases}
		\end{equation}

In the transverse channel, the amplitudes valid for generic values of $R$ read
		\begin{equation}
			\begin{aligned}
				\Tilde{\mathcal{A}}= 2^{-5} 2v & \left \{ \left [ \left ( n_1 + n_2 + n_3 + n_4 \right )^2 W_{4n} + \left ( n_1 - n_2 + n_3 - n_4 \right )^2 W_{4n+2} \right ]  V_8  \right. \\
				& - \left.\left [ \left ( n_1 + n_2 - n_3 - n_4 \right )^2 W_{4n} + \left ( n_1 - n_2 - n_3 + n_4 \right )^2 W_{4n+2} \right ] S_8 \right \} \ , 
			\end{aligned}
		\end{equation} 
and
		\begin{equation}
				\Tilde{\mathcal{M}}= - 2 \cdot 2 v  \left \{ \left ( n_1 + n_2 +n_3 + n_4 \right )  \hat{V}_8 \ W_{4n} - \left ( n_1 - n_2 - n_3 + n_4 \right )\hat{S}_8 \ W_{4n+2} \right \}\ ,
		\end{equation}
For rational values of the radius, the windings sum reduce to eq. \eqref{wsumratMth}.

\end{appendix}

\end{document}